# Surface science using coupled cluster theory via local Wannier functions and in-RPA-embedding: the case of water on graphitic carbon nitride

Tobias Schäfer,[1,*] Alejandro Gallo,[1] Andreas Irmler,[1] Felix Hummel,[1] and Andreas Grüneis[1]

[1] *Institute for Theoretical Physics, TU Wien, Wiedner Hauptstraße 8-10/136, A-1040 Vienna, Austria*

A first-principles study of the adsorption of a single water molecule on a layer of graphitic carbon nitride employing an embedding approach is presented. The embedding approach involves an algorithm to obtain localized Wannier orbitals of various types expanded in a plane-wave basis and intrinsic atomic orbital projectors. The localized occupied orbitals are employed in combination with unoccupied natural orbitals to perform many-electron perturbation theory calculations of local fragments. The fragments are comprised of a set of localized orbitals close to the adsorbed water molecule. Although the surface model contains more than 100 atoms in the simulation cell, the employed fragments are small enough to allow for calculations using high-level theories up to the coupled cluster ansatz with single, double and perturbative triple particle–hole excitation operators (CCSD(T)). To correct for the missing long-range correlation energy contributions to the adsorption energy, we embed CCSD(T) theory into the direct random phase approximation, yielding rapidly convergent adsorption energies with respect to the fragment size. Convergence of computed binding energies with respect to the virtual orbital basis set is achieved employing a number of recently developed techniques. Moreover, we discuss fragment size convergence for a range of approximate many-electron perturbation theories. The obtained benchmark results are compared to a number of density functional calculations.

## I. INTRODUCTION

First-principles modeling plays a pivotal role in the understanding of surface phenomena at the atomistic level. Adsorption processes, for instance, can only be understood with a quantum mechanical theory based on the many-electron Schrödinger equation. Density functional theory (DFT) calculations in the Kohn-Sham framework of approximate exchange and correlation (xc) energy functionals are the most popular approach for many theoretical surface science studies due to their good trade-off between accuracy and computational cost. However, the presence of uncontrolled approximations in the employed xc functionals often make it difficult to achieve the desired level of accuracy. Alternative methods, such as many-electron perturbation theories, achieve in many systems a more controllable accuracy and could complement DFT calculations favorably as long as the number of particles does not become computationally intractable. However, despite the fact that molecular adsorption can often be understood as a relatively local process, it is challenging to take full advantage of locality and realize a computational model that requires a quantum mechanical description of only a small number of particles in the vicinity of the adsorption site without compromising accuracy. In this work we apply a recently presented quantum mechanical embedding approach based on a plane-wave basis set implementation to a prototypical molecular adsorption problem. We obtain highly accurate adsorption energies on the level of quantum chemical many-electron theories up to coupled cluster with single, double and perturbative triple particle–hole excitation operators (CCSD(T)).

Here, we investigate the case of the adsorption of a single water molecule on graphitic carbon nitride ($gC_3N_4$). Graphitic carbon nitride is a promising compound for catalysis and energy storage[1–4]. It exhibits a high thermal stability and is easily processable from abundant materials. In addition, its band gap is sufficiently large to overcome the endothermic character of the water-splitting reaction pathway [5], which makes it interesting for applications in photocatalysis. This has also motivated theoretical studies of molecular adsorption processes [6, 7]. An initial step in such a reaction pathway is the adsorption of a single water molecule on the surface. Theoretical studies can provide estimates of the adsorption energy but have so far been limited to calculations on the level of DFT. This can partly be attributed to the fact that the two dimensional surface is corrugated and needs to be modeled using relatively big supercells, making it difficult to apply computationally more expensive but accurate methods, which are typically limited to unit cells containing only a few ten atoms. Therefore it can be assumed that embedding approaches, as described in the present work, help to remove limitations imposed by the simulation cell.

The theoretical study of the adsorption process of a single water molecule on a two dimensional material unveils strengths and weaknesses of many widely-used electronic structure theories. The importance of weak van der Waals (vdW) interactions in such systems can not be overestimated. Unfortunately, the ambiguous choice of the xc functional in DFT calculations and the availability of numerous approximations that aim at accounting for the effect of vdW interactions can make it difficult to achieve a faithful *ab initio* prediction of adsorption energies. Benchmark results can provide helpful insight into which functional is the most suitable for a particular system. Although high level methods such as CCSD(T) or quantum Monte Carlo (QMC) theories have successfully produced accurate benchmark results for periodic systems, the studied surface models often have been limited to a few ten atoms only [8–10]. Here, we advance our recently published embedding strategy [11] which enables a systematic path to approach the accuracy of CCSD(T) for regional phenomena in large systems of more than thousand electrons or periodic supercells with edge lengths of more than 20Å.

* tobias.schaefer@tuwien.ac.at

Regional embedding techniques for first-principles methods, in general, form an active research field [12–21]. This strategy was likewise adapted for many-electron correlation methods applied to molecules as well as periodic systems [11, 22–36]. However, regional phenomena in extended (periodic) environments exhibit long-range correlation effects, such as lattice relaxations or electron dispersion, which often exceed the size of a local fragment. As we show in this manuscript, the bonding of the water molecule to $gC_3N_4$, cannot exclusively be explained by H-N hydrogen bonds but also long-range van der Waals interactions play a considerable role. Additionally, ionic relaxations alter the electron density far away from the adsorption site. Hence, restricting electron correlation to local fragments only introduces sizable errors. A proper treatment of electron correlation between the local fragment and the environment is therefore indispensable in surface science and other regional phenomena, such as defect or polaron formation in solids [11, 35]. To achieve this, our plane-wave based scheme seamlessly embeds a high-level into a low-level correlation method. A convergence of the adsorption energy with respect to the local fragment size corresponds to a systematic approach to the high-level accuracy for the full system. Rapid convergence can be achieved if the low-level and high-level dispersions are quantitatively similar, as is observed for embedding coupled cluster theory (CC) into the direct random phase approximation (RPA) [11]. The choice of the RPA as a low-level correlation method is particularly favorable since it can be evaluated efficiently with cubic scaling algorithms using a plane-wave basis [37].

The paper is organized as follows. In section II we outline our scheme to embed a high-level into a low-level correlation method in detail, including the algorithm to construct localized Wannier orbitals as well as technical details to the many-electron correlation methods. The performance and results of our embedding scheme applied to the adsorption energy of $H_2O$ on $gC_3N_4$ are presented and discussed in section III A and III B, respectively.

## II. METHODS AND IMPLEMENTATIONS

We model the adsorption by means of a $3 \times 3$ single-layer of graphitic carbon nitride ($gC_3N_4$) in a supercell with periodic boundary conditions. We consider the free

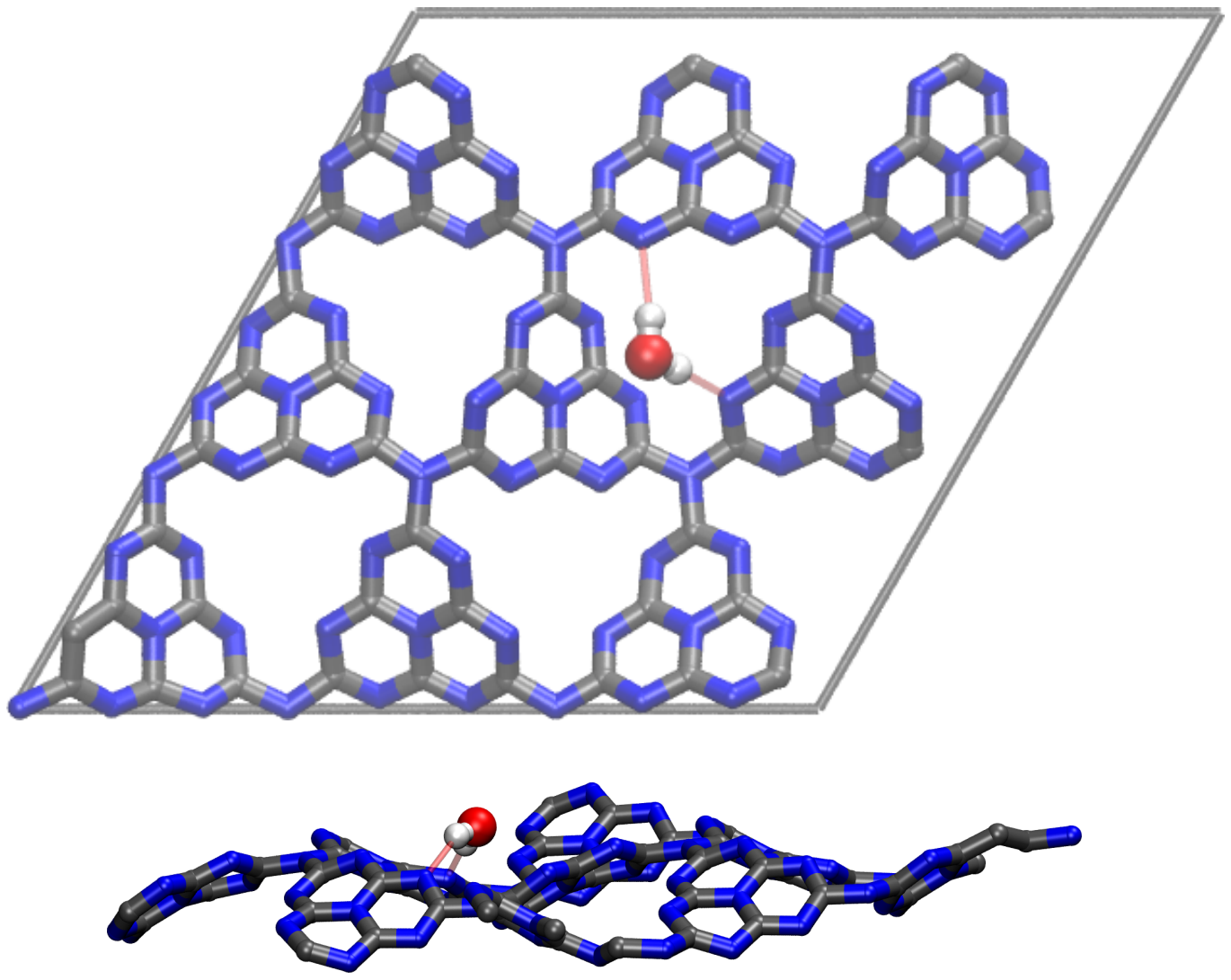

Figure 1. Top and side view of the periodic $gC_3N_4$supercell model with the adsorbed water molecule. The carbon atoms and covalent bonds are displayed in gray, and nitrogen in blue. Hydrogen bonds are indicated in light red. The cell edges are sketched in the $xy$-plane for the top view, while the height $z$ is chosen as 20 Å but is not shown.

energy (equivalently, the ground state energy) difference

$$E_{\mathrm{ad}} = E(\mathrm{H_2O@gC_3N_4}) - E(\mathrm{gC_3N_4}) - E(\mathrm{H_2O}) \, , \quad (1)$$

where three equally shaped supercells are employed, containing both the adsorbed water on the layer ($H_2O@gC_3N_4$), the layer alone ($gC_3N_4$), and the molecule alone ($H_2O$). The supercell for $H_2O@gC_3N_4$ contains 584 valence electrons from 54 carbon, 72 nitrogen atoms and the water molecule using a vacuum of 20 Å, as depicted in Fig. 1. This corresponds to 292 spin-restricted occupied Bloch bands. The equilibrium positions of the atoms are obtained from a ionic relaxation using the DFT-PBE functional for all three unit cells and can be found in the supplementary material [38]. The computationally most expensive steps of these free energy calculations arise from the many-electron correlation calculations, which are treated in an efficient embedding framework. We explain this embedding procedure in detail in Sec. II A.

All calculations start from a periodic spin-restricted Hartree-Fock (HF) reference of the entire supercell at the $\mathbf{\Gamma}$-point. We use the plane-wave based Vienna Ab-Initio Simulation Package (VASP) [39] to calculate the HF ground state, natural orbitals [40, 41], as well as the many-electron correlation at the second-order Møller-Plesset perturbation theory (MP2) [42, 43] and RPA [37] level of theory. The basis set is defined by the single cut-off parameter `ENCUT` = 500 eV and leads to about 183 000 plane-wave basis vectors which corresponds to more than 600 basis vectors per occupied band. VASP makes use of the projector augmented wave (PAW) method [44] in combination with frozen core potentials, see Tab. I for details. The coupled cluster calculations, coupled cluster with singles and doubles (CCSD) and CCSD(T), are performed with the `cc4s` code [10, 45]. All levels of many-electron correlation calculations are based on the exact same set of two-electron integrals, avoiding implementation-specific numerical issues. The calculations are performed on the Vienna Scientific Cluster equipped with 48 Intel Platinum 8174 processors and 384 GB per node.

Table I. Valence electrons, PAW cutoff radii $r_C$, and the default plane-wave cutoff `ENMAX` of the used PAW potentials. For all calculations a cutoff of `ENCUT` $= 500\,\text{eV}$ was used.

| Element | Valence | $r_C$ (Å) | `ENMAX` (eV) |
|---|---|---|---|
| H | $1s^1$ | 1.10 | 300.000 |
| C | $2s^2\,2p^2$ | 1.60 | 413.992 |
| N | $2s^2\,2p^3$ | 1.60 | 420.902 |
| O | $2s^2\,2p^4$ | 1.60 | 434.431 |

### A. Embedding scheme

Our embedding scheme allows the controllable regional embedding of a high-level into a low-level correlation method for calculations in large supercells. To this end we partition the electron density of the entire supercell into two fragments, a local region of interest (F) and the rest of the cell (R). In the considered adsorption process, the fragment F consists of the water molecule (in the case of the $H_2O@gC_3N_4$ supercell) and nearby parts of the densities of the $gC_3N_4$ layer. The construction of the fragments is achieved by selecting localized Wannier orbitals. In the following, we restate the individual technical steps of our embedding scheme as presented in our previous work [11]. The main difference consists in a new implementation of an algorithm for the construction of localized Wannier functions of various types for periodic systems in the plane-wave basis, which we discuss in detail in Sec. II B.

Once the density of the fragment F and the rest R are defined by two distinct sets of localized Wannier functions, the correlation energy calculation using method M can be restricted to orbitals in F or R (see six steps below), denoted as $E_{\mathrm{F}}^{\mathrm{M}}$ or $E_{\mathrm{R}}^{\mathrm{M}}$, respectively. We can then decompose the correlation energy of the entire cell as

$$E^{\mathrm{M}} = E_{\mathrm{F}}^{\mathrm{M}} + E_{\mathrm{R}}^{\mathrm{M}} + E_{\mathrm{F}\leftrightarrow\mathrm{R}}^{\mathrm{M}} \,. \tag{2}$$

This decomposition is exact since $E_{\mathrm{F}\leftrightarrow\mathrm{R}}^{\mathrm{M}}$ is implicitly defined by the above equation, while all other quantities are explicitly defined. The contribution $E_{\mathrm{F}\leftrightarrow\mathrm{R}}^{\mathrm{M}}$ can be interpreted as the inter-fragment correlation. Embedding the high-level method $\mathrm{M}_1$ into the low-level method $\mathrm{M}_2$ (in short $\mathrm{M}_1$:$\mathrm{M}_2$) can then intuitively be defined by

$$\begin{aligned} E^{\mathrm{M}_1:\mathrm{M}_2} &= E_{\mathrm{F}}^{\mathrm{M}_1} + E_{\mathrm{R}}^{\mathrm{M}_2} + E_{\mathrm{F}\leftrightarrow\mathrm{R}}^{\mathrm{M}_2} \,, \\ &= E^{\mathrm{M}_2} + \left(E_{\mathrm{F}}^{\mathrm{M}_1} - E_{\mathrm{F}}^{\mathrm{M}_2}\right) . \end{aligned} \tag{3}$$

Provided the low-level method $\mathrm{M}_2$ can be applied to all electrons in the entire supercell, the presented embedding scheme accounts for all many-electron correlation effects, but at mixed levels of theory.

The actual restriction of a correlation method to a local fragment of the density is by no means trivial nor unique. The following six steps explain our approach which involves an efficient compression of the unoccupied space and allows for a systematic convergence of the correlation energy with respect to the size of the fragment F.

(i) *Hartree-Fock ground state*
Self-consistent calculation of the $N_{\mathrm{occ}}$ occupied HF Bloch orbitals.

(ii) *Unoccupied space*
Diagonalize the HF Hamiltonian in the full plane-wave basis to obtain all $N_{\mathrm{vir}}$ unoccupied Bloch orbitals.

(iii) *First compression of unoccupied space*
Calculate the natural orbitals at the level of the RPA [41] for the entire supercell. Rediagonalize the Fock matrix using the first $N_{\mathrm{vir}}^{\mathrm{pre}} = 60 \cdot N_{\mathrm{occ}}$ natural orbitals (NOs) in order to precompress the unoccupied space by a factor of ten as compared to the full plane-wave basis.

(iv) *Build fragments*
Transform all occupied Bloch orbitals to $N_{\mathrm{occ}}$ Wannier orbitals. Here we use the intrinsic bonding orbitals (IBOs) which are a special case of the Pipek-Mezey (PM) orbitals, see Sec. II B. To each atom we assign a set of orbitals based on the partial charge of the Wannier orbitals. The fragment is then defined by selecting atoms in the $gC_3N_4$ layer, based on the radial distance from the adsorbed water molecule, which is included in each fragment. Finally, we diagonalize the HF Hamiltonian in the basis of the $N_{\mathrm{occ}}^{\mathrm{loc}}$ selected Wannier orbitals (re-canonicalization).

(v) *Second compression of unoccupied space*
Calculate the natural orbitals of the fragment only, based on an approximated MP2 scheme [40] using $N_{\mathrm{vir}}^{\mathrm{pre}}$ pre-compressed orbitals to further compress the unoccupied space to $N_{\mathrm{vir}}^{\mathrm{loc}} = Y \cdot N_{\mathrm{occ}}^{\mathrm{loc}}$ orbitals. For MP2 and RPA we choose $Y = 60$ and for CCSD and CCSD(T) we choose $Y = 20$ augmented by a basis set correction procedure.

(vi) *Local correlation energy*
Decompose the electron repulsion integrals (ERI) into auxiliary three index quantities using $N_{\mathrm{F}}$ auxiliary field variables, i.e. auxiliary basis functions [11, 45]) Calculate the correlation energy (MP2, RPA, CCSD, CCSD(T), ...) for $N_{\mathrm{occ}}^{\mathrm{loc}}$ local occupied and $N_{\mathrm{vir}}^{\mathrm{loc}}$ unoccupied orbitals and employing $N_{\mathrm{F}}$ auxiliary field variables.

Please note, that steps (i) and (ii) are common preparatory steps for regular correlation energy calculations. The embedding procedure mainly consists in steps (i)+(ii)+(iv)+(vi). The compression of the unoccupied space in steps (iii) and (v) primarily serve to increase the computational efficiency. Further technical details to the correlation methods are outlined in Sec. II C

### B. Localized Wannier functions

Here, we present the theory and implementation of the algorithm to construct localized Wannier functions of various types for large supercells in the Vienna Ab Initio Simulation Package VASP. The algorithm essentially follows the work of Abrudan et al. [46] which has already been used for orbital localization of molecules and solids in Refs. [47–49]. Localized Wannier orbitals $|\mathcal{W}_i\rangle$ are obtained by a unitary transformation of the (delocalized) spin-orbitals $|\chi_j\rangle$,

$$|\mathcal{W}_j\rangle = \sum_i^{N_{\text{occ}}} u_{ij}\,|\chi_i\rangle\ . \tag{4}$$

The expected Brillouin zone integration [50] is neglected here, since we currently aim for large supercells using a Γ-point-only sampling. The unitary matrix $u_{ij}$ is determined by the optimization (maximization/minimization) of a cost functional $\mathcal{L}$ which ultimately defines the localized Wannier orbitals. For instance, the PM localization [51] is defined by maximizing the atomic partial charges,

$$\mathcal{L}^{\text{PM}}[\{u_{ij}\}] = \sum_i^{N_{\text{occ}}} \sum_A^{N_{\text{atoms}}} |\,\langle \mathcal{W}_i|\boldsymbol{P}_A|\mathcal{W}_i\rangle\,|^p\ , \tag{5}$$

where $p \geq 2$ and $\boldsymbol{P}_A = \sum_{\mu\in A}|\mu\rangle\langle\mu|$ is a projector onto atom centered functions $|\mu\rangle$ at atom $A$. In the case of spin-polarization, a separate cost functional is considered for each spin. While our implementation allows the optimization of any cost functional $\mathcal{L}$ which is polynomial in $u_{ij}$ (including also other optimizations beside orbital localization), we choose the PM localization cost functional and its variants as a specific example. In particular, we consider the IBOs which are obtained by choosing intrinsic atomic orbitals (IAOs) $|\mu^{\text{IAO}}\rangle$ for the projector $\boldsymbol{P}_A$ in the construction of the PM cost functional [52].

The algorithm to maximize/minimize the considered cost functional is essentially an iterative unitary matrix constrained optimization (UMCO) procedure. Starting from a stochastic unitary matrix $\boldsymbol{U} = (u_{ij})$ each iteration step propagates the unitary matrix according to

$$\boldsymbol{U} \longrightarrow \exp(\pm\mu\boldsymbol{G})\,\boldsymbol{U}\ . \tag{6}$$

The Riemannian derivative $\boldsymbol{G} = \boldsymbol{\Gamma}\boldsymbol{U}^\dagger - \boldsymbol{U}\boldsymbol{\Gamma}^\dagger$ is anti-Hermitian, $\boldsymbol{G}^\dagger = -\boldsymbol{G}$, and is calculated from the Euclidean derivative $\boldsymbol{\Gamma} = \partial\mathcal{L}/\partial\boldsymbol{U}^*$. A conjugate-gradient technique based on the Polak-Ribière-Polyak update factor is applied in order to speed-up the convergence. We refer the reader to Ref. [46] for details. The iterations are considered as converged, once the norm obeys $\langle\boldsymbol{G},\boldsymbol{G}\rangle = \text{tr}(\boldsymbol{G}\boldsymbol{G}^\dagger)/2 < \delta$ where we choose $\delta = 10^{-10}$. For the case of the PM functional we find

$$\begin{aligned}\Gamma_{ij}^{\text{PM}} &= \frac{\partial\mathcal{L}^{\text{PM}}}{\partial u_{ij}^*} \\ &= p\sum_A^{N_{\text{atoms}}}\sum_k^{N_{\text{occ}}}\langle\chi_i|\boldsymbol{P}_A|\chi_k\rangle\,u_{kj}\left|\sum_{lm}^{N_{\text{occ}}}\langle\chi_l|\boldsymbol{P}_A|\chi_m\rangle\,u_{lj}^*\,u_{mj}\right|^{p-1}\end{aligned} \tag{7}$$

The optimal step size $\mu$ obeys $\frac{\partial}{\partial\mu}\mathcal{L}[\exp(\pm\mu\boldsymbol{G})\boldsymbol{U}] = 0$ and is estimated by a polynomial line search approximation as described in Ref. [46]. A positive sign $+\mu$ is used for the maximization (e.g. for the PM functional), while $-\mu$ corresponds to a minimization of $\mathcal{L}$.

The plane-wave basis and the specifics of the PAW method have to be respected only in the preparatory calculation of the overlaps (scalar product) between $\langle\chi_i|\mu\rangle$ to compute the action of the atomic projectors $\boldsymbol{P}_A$. The calculation of these overlaps are parallelized and has previously been implemented for an interface between VASPand wannier90 in Ref. [53]. In contrast, the iterations (6) are not parallelized, instead run individually on each rank but with separate stochastic initialization. This redundancy allows for a simple consistency check of the reached extremum of the cost functional in order to reveal "local minimum" issues, if present. The computational complexity of the iterations are dominated by matrix multiplications, hence follow a cubic scaling in the number of orbitals to localize. Timings are presented in Sec. III. The cubic scaling of the evaluation of $\Gamma_{ij}^{\text{PM}}$ from Eq. (7) is not obvious and can be achieved using the following equivalent formulation,

$$\Gamma_{ij}^{\text{PM}} = p\sum_A^{N_{\text{atoms}}}(\boldsymbol{X}_A\boldsymbol{Y}_A)_{ij}\left|(\boldsymbol{Y}_A^\dagger\boldsymbol{Y}_A)_{jj}\right|^{p-1}\ . \tag{8}$$

The atomic overlap matrix $(\boldsymbol{X}_A)_{i\mu} = \langle\chi_i|\mu\rangle$ with $\mu$ centered on atom $A$ only and its unitary rotated variant $\boldsymbol{Y}_A = \boldsymbol{X}_A^\dagger\boldsymbol{U}$ was used. For each atom, matrix multiplications have to be performed, resulting in a complexity of $N_{\text{atoms}}\cdot N_{\text{occ}}^2(N_\mu/N_{\text{atoms}}) = N_{\text{occ}}^2 N_\mu$, where $N_\mu/N_{\text{atoms}}$ is the number of atom centered functions per atom.

As atom centered functions we choose IAOs, $|\mu\rangle \rightarrow |\mu^{\text{IAO}}\rangle$, which can be obtained from atomic functions $|f_\mu\rangle$ by a simple projection procedure,

$$|\mu^{\text{IAO}}\rangle = (\mathbb{1} + \mathcal{O} - \widetilde{\mathcal{O}})|f_\mu\rangle\ , \tag{9}$$

where we follow the formulation of Janowski [54]. The atomic functions $|f_\mu\rangle$ may be HF or DFT solutions of an isolated atom, but also Hydrogen-type or Slater-type functions are a valid choice. The two projectors are defined by

$$\mathcal{O} = \sum_i^{N_{\text{occ}}}|\chi_i\rangle\langle\chi_i|\ ,\quad \widetilde{\mathcal{O}} = \sum_i^{N_{\text{occ}}}|\widetilde{\chi}_i\rangle\langle\widetilde{\chi}_i|\ . \tag{10}$$

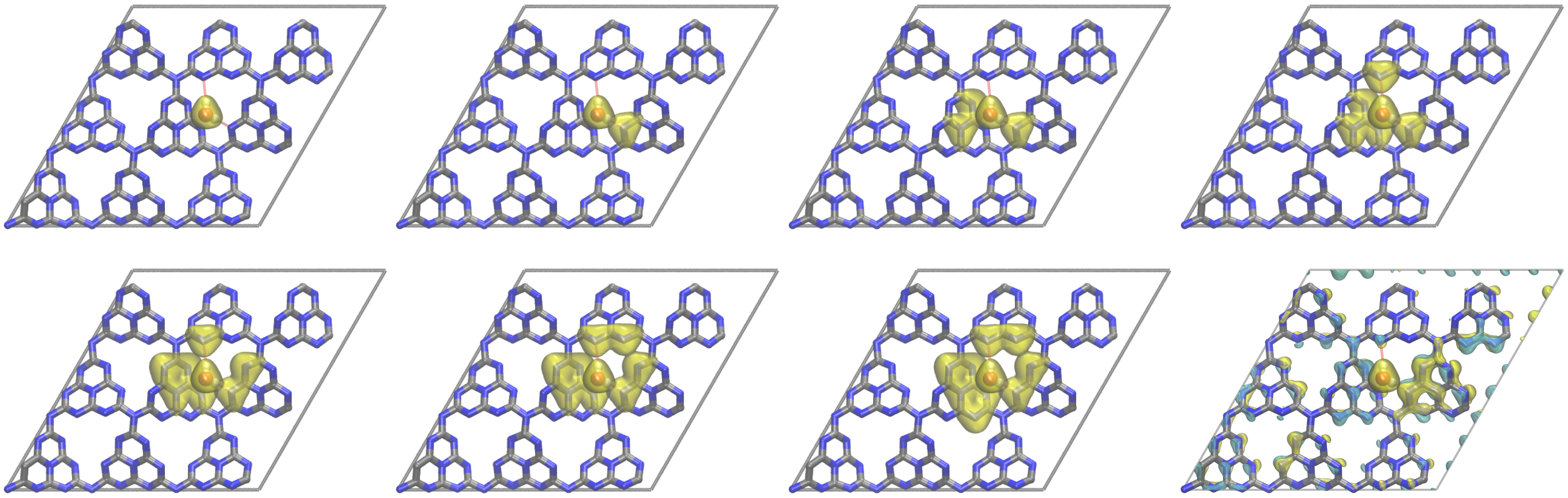

Figure 2. From left to right and from top to bottom: isosurfaces of the densities of the fragments 1 to 7. These fragments correlate 4, 8, 16, 24, 32, 36, and 40 occupied orbitals. Bottom right: Change in the electron density between the ionically relaxed layer with adsorbed water molecule and the pristine layer. Positive and negative change in the density is illustrated in yellow and blue, respectively.

While $\mathcal{O}$ is simply the projection onto the occupied space, $\widetilde{\mathcal{O}}$ projects onto the space spanned by

$$|\widetilde{\chi}_i\rangle = \text{orth}\left[\sum_{\mu\nu} |f_\mu\rangle\, S^{-1}_{\mu\nu}\, \langle f_\nu|\chi_i\rangle\right] \tag{11}$$

with the overlap matrix $S_{\mu\nu} = \langle f_\mu|f_\nu\rangle$. The subsequent orthogonalization is denoted as "orth". The orbitals decorated with a tilde, $|\widetilde{\chi}_i\rangle$, can be understood as the occupied orbitals as described by the certainly incomplete "minimal basis" spanned by $\{|f_\mu\rangle\}$. Note that the occupied orbitals $|\chi_i\rangle$ originate from a preceding mean-field calculation using the plane-wave basis, which we consider as the complete basis. The action of $\mathcal{O}-\widetilde{\mathcal{O}}$ on $|f_\mu\rangle$ in Eq. (9) constructs the necessary augmentation, such that the IAOs $|\mu^{\text{IAO}}\rangle$ form an exact basis of the occupied space as described by the complete basis. This is true as long as no single occupied orbital is orthogonal to the atomic functions, $\sum_\nu \langle f_\nu|\chi_i\rangle \neq 0$, $\forall i$. Additionally, if the density spillage of the tilted orbitals is not too large, $|\mu^{\text{IAO}}\rangle$ exhibits almost the same spatial shape as the atomic function $|f_\mu\rangle$. In this case, the IAOs can be considered as a minimal but exact atom centered and local basis for the occupied space. Furthermore, the atomic partial charge in the PM cost functional (5) are stable against different choices of atomic functions $|f_\mu\rangle$ if the IAOs are used for the projectors $\boldsymbol{P}_A$, as shown in Ref. [52]. The resulting Wannier orbitals are thus the IBOs. The exponential decay of the gradient norm $\langle \boldsymbol{G}, \boldsymbol{G}\rangle$ using the presented UMCO algorithm applied to water on $gC_3N_4$ is shown in Fig. 3.

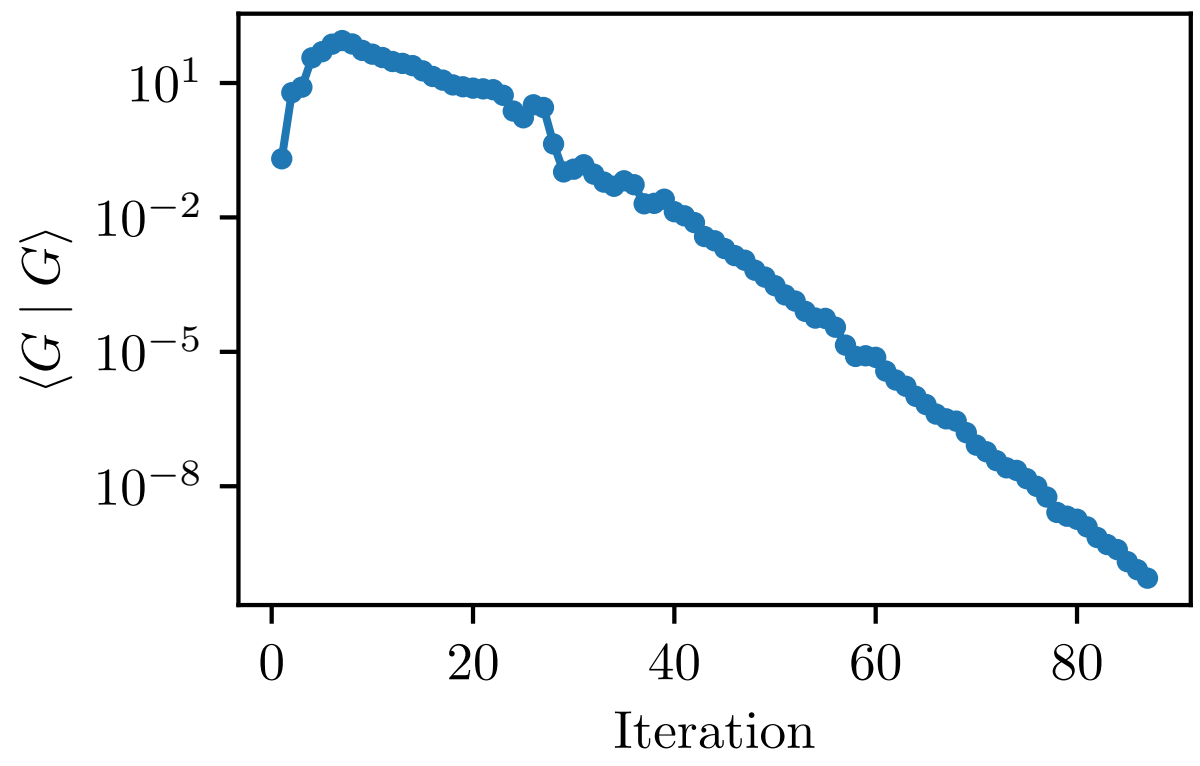

Figure 3. Convergence of the iteration steps in the UMCO localization scheme to construct IBOs for the supercell of $H_2O@gC_3N_4$. The gradient norm decays exponentially and reaches $10^{-10}$ after 87 iteration steps. Each iteration step took about 0.46 seconds on one single core and rotates all 292 orbitals at once, as formulated by Eq. (6).

### C. Electron correlation methods

For the post-HF calculations of the localized electron states in the fragment we employ MP2, RPA, CCSD, and CCSD(T) theory. The localized electronic states are re-canonicalized using VASP, making it possible to employ canonical formulations of all many-electron theories.

The virtual orbitals are computed using the following two-step procedure as already outlined in Ref.[11]. We first compute the RPA natural orbitals (RPANOs) for the full supercell prior to the localization procedure of the occupied orbitals[41]. We truncate the RPANOs to a subset constructed from 60 natural orbitals per occupied spatial orbital and re-canonicalize the Fock matrix in this subspace. We refer to the corresponding total number of unoccupied orbitals at this stage by $N_{\text{vir}}^{\text{pre}}$. In the second compression step we compute approximate MP2 natural orbitals for a set of localized occupied orbitals. To this end we calculate the virtual-virtual ($N_{\text{vir}}^{\text{pre}} \times N_{\text{vir}}^{\text{pre}}$) block of an approximate MP2 reduced density matrix defined by Eq.2 in Ref. [40]. In this step, the sum over the occupied orbitals for the expression of the reduced density matrix is restricted to a set of semi-canonicalized set of occupied orbitals that are localized on the fragment. As a consequence, the resulting approximate MP2NOs with the largest occupation numbers are also localized on the fragment. We refer to the corresponding number of employed NOs by $N_{\rm vir}^{\rm loc}$.

Once an optimized set of localized occupied and virtual orbitals has been obtained for each fragment, the required information for the post-HF calculations is passed to the `cc4s` code, which performs MP2, CCSD and CCSD(T) calculations. In addition to the HF orbital energies, the decomposed ERI are required for the MP2 and CC calculations. We note that certain integrals are computed on-the-fly in order to minimize the memory footprint of the CCSD calculations. For the decomposed ERI, we employ a procedure that was previously described in Ref. [45] and involves auxiliary three index quantities using $N_{\rm F}$ optimized auxiliary basis functions. All Coulomb integrals, $V_{sr}^{pq}$, needed by coupled cluster theories can be computed using the following expression

$$V_{sr}^{pq} = \sum_{F=1}^{N_F} \Gamma^{*pF}_{s} \Gamma^{q}_{rF}, \tag{12}$$

where $p, q, r$ and $s$ refer to (occupied or virtual) orbital indices. $F$ refers to auxiliary basis functions that have been obtained using a singular value decomposition outlined in Ref. [45]. For the present work, due to the local character of both occupied and virtual orbitals, significant reductions of the auxiliary basis set size are possible without compromising the precision of computed correlation energies. We have carefully checked that the computed correlation energies are converged to within meV with respect to the size of the optimized auxiliary basis set. We stress that as a result of the optimisation procedure described above, all dimensions that contribute to the scaling of the computational complexity of the electron correlation methods exhibit a linear scaling with fragment size: $N_{\rm F} \propto N_{vir}^{\rm loc} \propto N_{\rm occ}^{\rm loc}$, despite using a plane-wave basis set.

To achieve basis set converged electronic correlation energy differences on all levels of theory we employ the following procedures. The MP2 correlation energy calculations are performed for each fragment using 60 natural orbitals per occupied orbital. In addition, an automatic basis set extrapolation to the complete basis set (CBS) limit is performed. The extrapolation procedure was previously described in Ref.[42]. Here, the extrapolation is performed for each electron pair correlation energy contribution to yield $\epsilon_{ij}^{\rm MP2}$. Throughout this work we refer to CBS limit estimates when referring to the MP2 energy.

The CCSD correlation energies are computed using 20 natural orbitals per occupied orbital for each fragment. To correct for the remaining basis set incompleteness error (BSIE) of CCSD correlation energies we employ a correction that was previously introduced in Ref. [55]. This correction includes contributions from two terms: (i) a BSIE correction on the level of MP2 theory and (ii) an approximation to the BSIE of the particle-particle-ladder (PPL) term contribution to the CCSD energy. We have concluded in Refs. [55–57] that these are indeed the dominant BSIE contributions on the level of CCSD theory for a wide range of properties. In passing we note that Refs. [58, 59] have shown related findings. The employed PPL correction is electron pair specific and depends on $\epsilon_{ij}^{\rm MP2}$. As described in Ref.[55], this approach (also referred to as FPb+Δps-ppl) yields CCSD reaction energies that deviate from the CBS limit by about 11 meV only, which is closer to the CBS limit than F12 calculations employing an aug-cc-pVTZ basis set. For the sake of brevity we will simply refer to the CCSD CBS limit energy estimates obtained in this manner as CCSD energies.

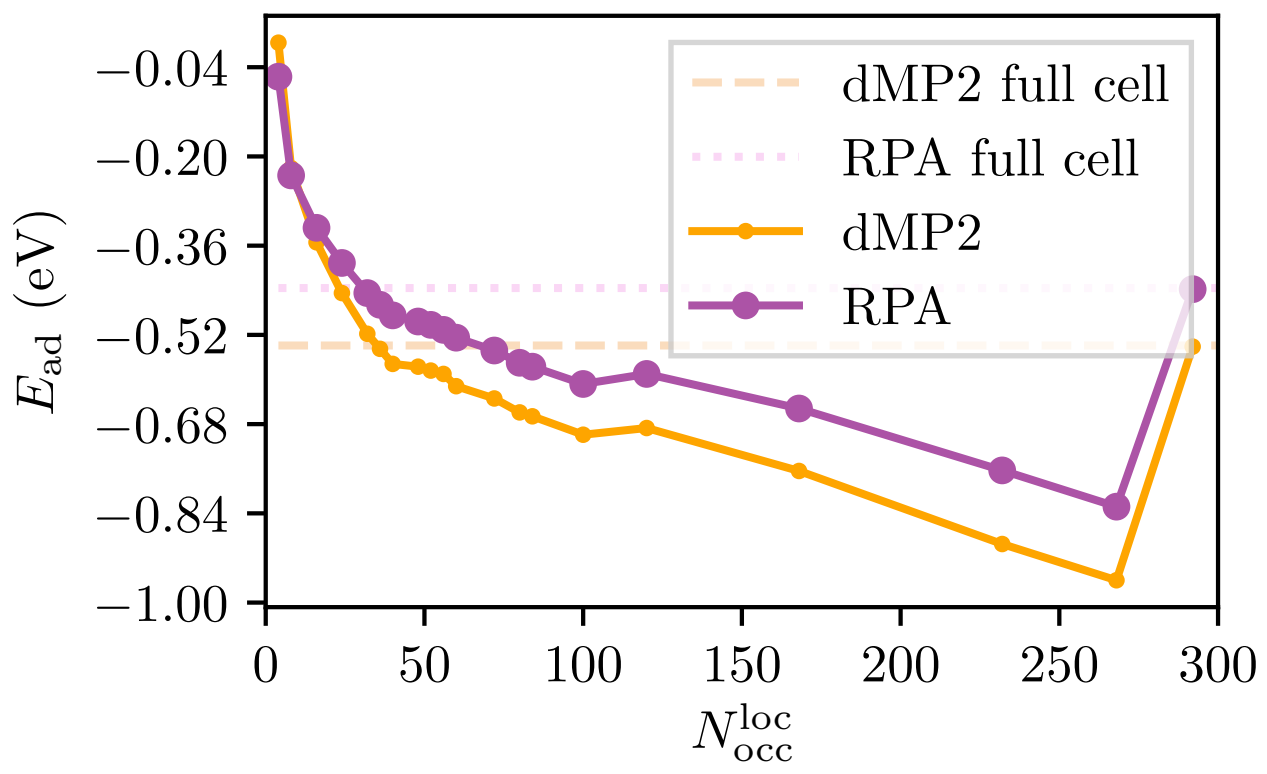

Figure 4. Non-embedded calculations of the adsorption energies from correlation restricted to the bare fragments at the level of dMP2 and RPA. Reference values for the full cell are also provided. Clearly, the bare fragments fail to approach the full cell result.

The (T) correlation energy contribution is computed using 20 natural orbitals per occupied orbital, which is considered a well converged CBS limit estimate. This is partly confirmed by benchmark results published in Ref.[55].

## III. RESULTS AND DISCUSSION

### A. Embedding approach

We now turn to the results obtained using the embedding approach and their convergence with respect to the number of occupied orbitals in the studied fragments. Figure 2 shows the first seven fragments built from selected IBOs as used for the embedding scheme. The localized occupied orbitals included in the employed fragments are selected according to their distance from the water molecule.

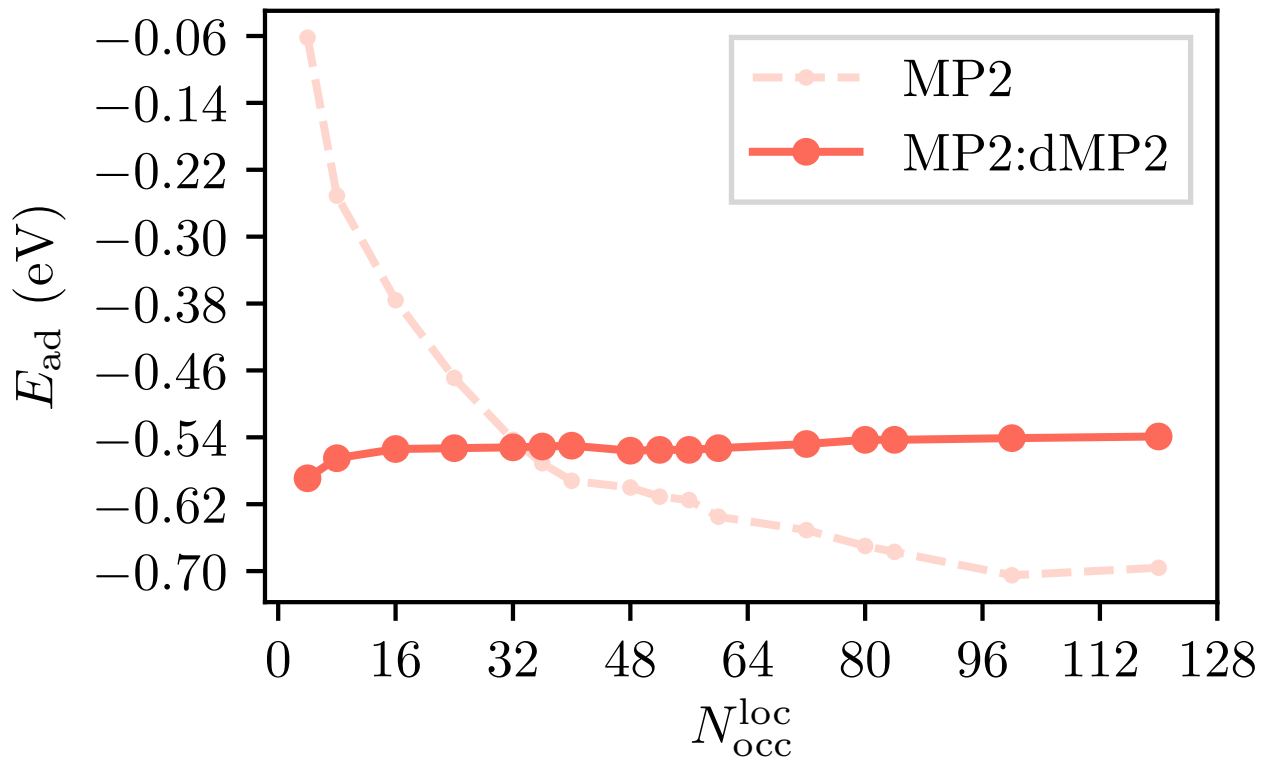

Figure 5. The adsorption energy as calculated by embedding different fragment sizes of MP2 into dMP2 as well as for the non-embedded MP2 fragments. A convergence is only visible for the embedded MP2:dMP2 correlation calculation.

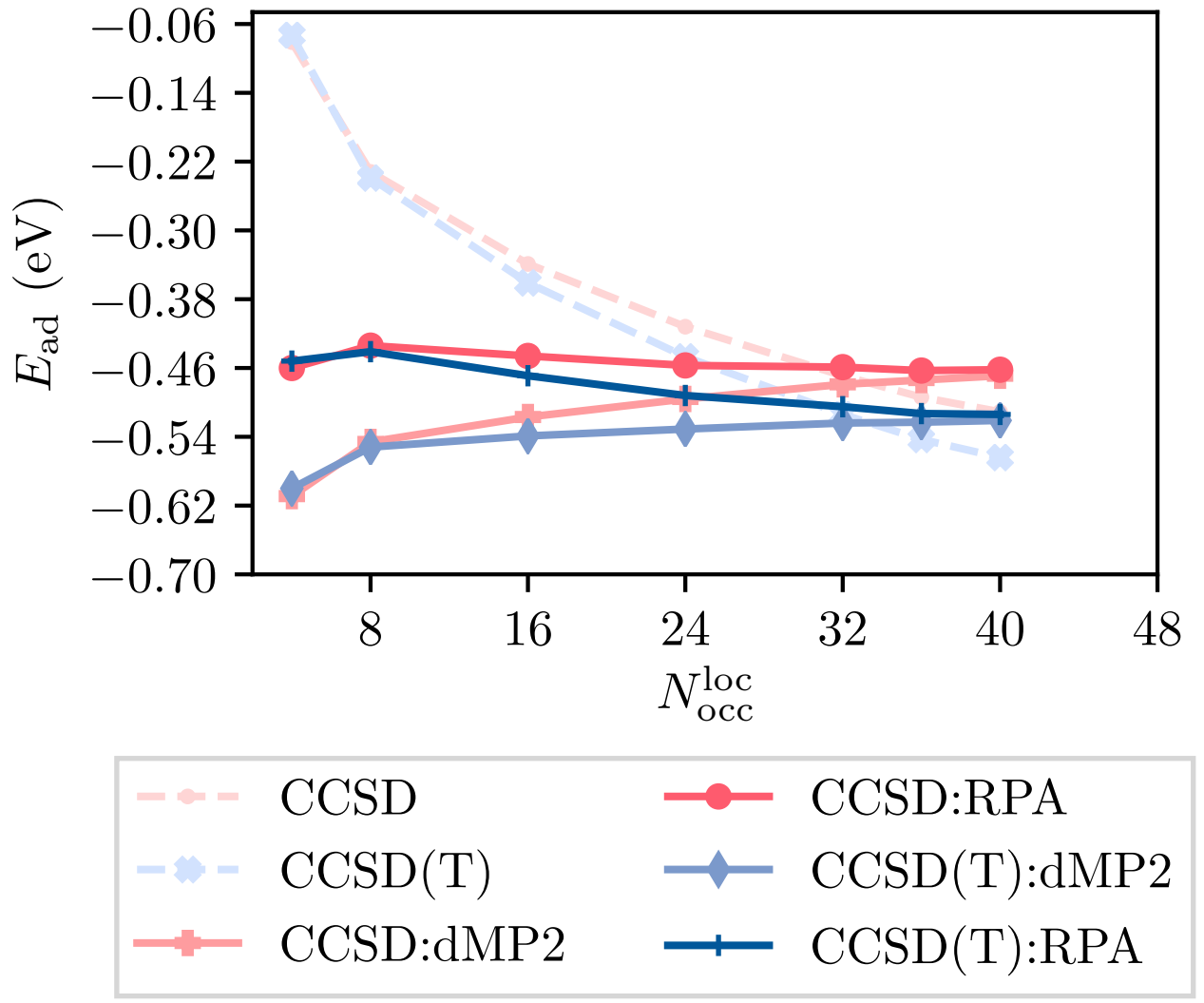

Figure 6. The adsorption energy as calculated by embedding different fragment sizes of CCSD and CCSD(T) into RPA and dMP2 as well as for the non-embedded fragments.

We first discuss the convergence of dMP2 and RPA adsorption energies with respect to fragment size. Fig. 4 depicts water adsorption energies retrieved as a function of the number of occupied orbitals in the fragment. The horizontal dashed lines show the dMP2 and RPA reference adsorption energies obtained from calculations of the full periodic cell. We note that the adsorption energies obtained for the fragments include electron correlation contributions of the bare fragments only, i.e. only $E_{\mathrm{F}}^{\mathrm{M}_1}$ from Eq. (3) without embedding into a low-level theory $\mathrm{M}_2$. As apparent from Fig. 4, no convergence with respect to the fragment size can be achieved for dMP2 and RPA in the present system for the considered fragments. This lack of convergence for the non-embedded fragments was already observed in Ref. [11] for the case of the ionically relaxed surface of $TiO_2$ adsorbing a water molecule. In contrast, the embedding into a low-level method, $\mathrm{M}_1$:$\mathrm{M}_2$, leads in all cases to a clearly smoother convergence. The MP2:dMP2 embedding rapidly approaches its limit at −0.54 eV, as shown in Fig. 5. The progression to fragments with up to 120 correlated bands reveals a remaining numerical noise of about 0.02 eV. We therefore estimate the error of our embedding results to this value.

Having observed that the embedding into a low-level method substantially improves the convergence, we now seek to better understand the source of the slow convergence of the non-embedded correlation energies in Fig. 4. Due to relaxations of the atoms caused by the water adsorption, the electron density is considerably altered in the entire supercell, as can be observed at the right bottom in Fig. 2. Note that the dipole of the water molecule alone would only alter the density in a much smaller region of the layer. The density change in the entire cell, however, diminishes the similarity of the fragments with and without the adsorbed water molecule, albeit the fragments were built using the same selection of atomic sites in the layer (step iv) of the corresponding supercells. The diminished similarity manifests itself in the fragments surfaces. As apparent from Fig. 2 this can not be eased using larger and larger fragments. Instead, the surface effect is only pushed along as the fragment size increases but is abruptly corrected as soon as the fragments reach the size of the supercell, i.e. when the surfaces melt into their periodic images. Conversely, the convergence of the non-embedded fragments is restored if the atomic sites are not relaxed, i.e. if the positions of the atoms in the layer are equivalent in the supercell with and without the adsorped water. We show the analog to Fig. 4 for the non-relaxed case in the supplementary material [38]. From this we conclude that the change in atomic positions in our studied material prohibit a local fragment-only approach to capture the electron correlation effects of the considered adsorption process.

In the following we discuss the performance of our embedding scheme for the coupled cluster methods. As mentioned above, we believe that embedding a high-level into a low-level correlation method allows to remedy two central issues of local fragments in periodic environments: (i) it accounts for the missing non-local long-range correlation and (ii) it cancels spurious long-range relaxation effects in the atomic positions. The latter is particularly important if lattice relaxations reduce the similarity between the considered supercells. Slightly displaced ions result in a change of the electron density, which can have an uncontrollable effect on the localized Wannier orbitals and accordingly on the surface character of the fragments. Nevertheless, even in situations without relaxation of the atomic positions, non-local long-range dispersion interactions can prohibit a fast and smooth convergence of the correlation energy with respect to the fragment size. In Ref. [11] we demonstrated that a rapid convergence can be observed if the low-level and high-level long-range dispersions are quantitatively similar. In the language of diagrammatic perturbation theory, we expect this if the low-level method is a diagramatic subset of the high-level theory, where the former can easily be restricted to capturing only long-range terms accurately. For the MP2:dMP2 and CCSD:RPA embedding this is confirmed by our presented results. In the first case, the low-level method accounts for the full long-range correlation, while the second case benefits from the fact that the direct ring diagrams capture important part of the screened dispersion correlation of CCSD. However, we have to assume equally that the long-range behavior of the perturbative triples in CCSD(T) cannot be well approximated by neither dMP2 nor RPA. The rapid convergence of CCSD(T):RPA and CCSD(T):dMP2 is therefore primarily due to the rapid decay of the triples themselves. Presumably, the seemingly faster convergence of CCSD(T):dMP2 can be attributed to the fortuitous proximity of the final energies of dMP2 and CCSD(T). Again, we believe that CCSD(T):RPA is generally more favorable since the underlying CCSD:RPA is guaranteed to converge quickly while the perturbative triples are hardly corrected.

Table II. The adsorption energy in units of meV of the water molecule as calculated by the different methods. The error is calculated as the difference to the CCSD(T):RPA result. Note that we explicitly write RPA@HF to avoid confusion with the DFT based RPA method, which is not considered in this work.

| Method | $E_{ad}$ | abs. error | %error |
|---|---|---|---|
| HF | −206 | +308 | -60 |
| PBE | −350 | +165 | -32 |
| PBE-D3 | −528 | -13 | +3 |
| PBE-TS | −550 | -35 | +7 |
| optB86B | −568 | -54 | +10 |
| PBE0-MBD | −537 | -22 | +4 |
| dMP2 | −539 | -25 | +5 |
| MP2:dMP2 | −539 | -25 | +5 |
| RPA@HF | −436 | +79 | -15 |
| CCSD:RPA | −462 | +53 | -10 |
| CCSD:dMP2 | −469 | +46 | -9 |
| CCSD(T):RPA | −514 | | |
| CCSD(T):dMP2 | −521 | | |

## B. Adsorption energies

We now discuss the converged adsorption energies obtained from each method considered. The results of each method are compiled in Tab. II. The HF result of −206 meV provides a reference to measure the contribution of the electron correlation of the water adsorption, since HF captures the full electrostatics but lacks electron correlation by definition. The PBE result indicates significant correlation effects, when comparing to the uncorrelated HF result. As corrected functionals we employed PBE-D3 (the method of Grimme et al. [60] with Becke-Jonson damping), PBE-TS (the Tkatchenko-Scheffler method [61]), optB86B (the method of Langreth and Lundqvist in the optimized form as presented in [62]), PBE0-MBD (the many-body dispersion energy method as presented in Ref. [63] using the scaling parameter $\beta = 0.85$). These corrections for long-range correlation suggest an even stronger impact, such that the total electron correlation contribution to the adsorption energy can be assumed to be of similar if not larger order of magnitude as the electrostatic energy. This suggestion is supported by the many-electron correlation methods dMP2 and RPA, which could be applied to the entire supercells. However, the large difference between dMP2 and RPA of about 100 meV prohibits a consistent picture. In passing we note that we perform RPA calculations using a HF reference, which is in contrast to the more prevalent approach of performing RPA calculations on top of a Kohn-Sham reference. The discrepancy between dMP2 and RPA comes as no surprise, since the accuracy of both methods is diminished in extended systems by the following systematic problems. To mention only the most eminent issues, the MP2 (as well as dMP2) method suffers from severe overestimations of the long-range correlation [64] due to the missing electron screening which, contrary, is considerably more accurately captured by the RPA, as observed for many materials [65–70]. Yet the RPA results may be deteriorated from the self-correlation problem of short-range correlation and from a too deep correlation hole [71–73]. The estimated CCSD(T) result finally confirms the well-known overcorrelation and undercorrelation trends of MP2 and RPA, respectively. With about 0.05 eV, the triples contribute 16% to the correlation contribution and 10% to the full adsorption energy. Having computed a reliable CCSD(T) adsorption energy, we can conclude that PBE-D3 and PBE0-MBD achieve an excellent trade-off between accuracy and cost for the present system. We note, however, that this is in contrast to a related study of water adsorption on a single layer of hexagonal boron nitride, where PBE-D3 and PBE0-MBD overestimated that adsorption energy compared to more accurate CCSD(T) and QMC findings[9].

## C. Timings

We also report the timings of the calculations in the steps (i)-(vi) from Sec. II A. All timings correspond to the largest system, i.e. $H_2O@gC_3N_4$. The HF iterations (step i) converged in about 210 core hours. The calculation of the unoccupied bands (step ii) takes about 3300 core hours. Natural orbitals at the RPA level (step iii) cost about 21000 core hours. The localization to construct the IBOs (step iv) finished in 0.2 core hours after 87 iteration steps. Timings for the calculation of the sec-ond compression (natural orbitals at the level of MP2, step v), as well as for step (vi), MP2 and CCSD(T), can be found in Fig. 7 for different fragment sizes.

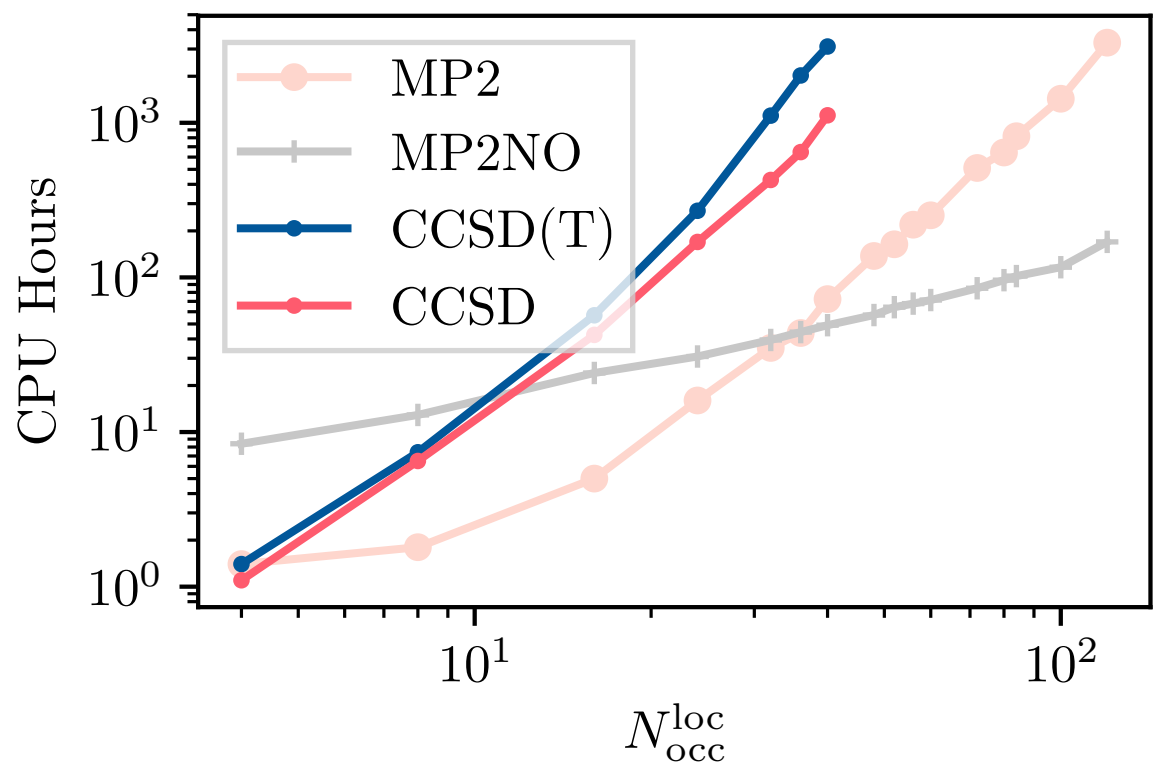

Figure 7. Log-log plot of the CPU hours for steps (v) and (vi) for various fragment sizes.

Apparently, the construction of the localized Wannier functions is the least expensive step.

## IV. CONCLUSION

We have calculated the adsorption energy of water on graphitic carbon nitride employing coupled cluster theory including singles doubles and perturbative triples, CCSD(T). Results at the level of MP2, RPA, and CCSD were also presented. This was made possible by an in-RPA-embedding that leverages the numerical similarity between CC and RPA for the electronic long-range correlation. Recently developed techniques were employed to reach the complete basis set limit for all correlation methods.

We furthermore presented the implementation of an algorithm that enables various orbital localizations for periodic systems, provided a proper cost functional can be defined. The construction of localized Wannier functions using the Pipek-Mezey cost functional (in the variant for intrinsic bonding orbitals) was demonstrated and used to define the local fragments for the embedding procedure.

Our findings show that fragment based techniques for black box correlation calculations of regional phenomena invariably need to correct for effects far outside the region of interest. These effects may be due to long-range dispersion correlation as well as to distorted local fragments caused by long-range density alterations which in turn are triggered by far reaching ionic relaxations of the regional process. We demonstrated that such effects are reliably captured by the RPA.

## SUPPLEMENTARY MATERIAL

See the supplementary material for the employed atomic structures (POSCAR files) as well adsorption energies without preceding ionic relaxations.

## ACKNOWLEDGMENTS

The authors thankfully acknowledge support and funding from the European Research Council (ERC) under the European Unions Horizon 2020 research and innovation program (Grant Agreement No 715594). T.S. extends his special thanks to Manuel Engel and Henrique Miranda for the enhanced parallelization efficiency to calculate the scalar products (overlaps) between the Bloch orbitals and local trial functions for the new UMCO module. We thank Livia Getzner for providing the relaxed atomic structure used in the present study. The computational results presented have been achieved in part using the Vienna Scientific Cluster (VSC).

## DATA AVAILABILITY

The data that supports the findings of this study are available within the article and its supplementary material [38].

---

[1] Arne Thomas, Anna Fischer, Frederic Goettmann, Markus Antonietti, Jens-Oliver Müller, Robert Schlögl, and Johan M. Carlsson, "Graphitic carbon nitride materials: variation of structure and morphology and their use as metal-free catalysts," J. Mater. Chem. **18**, 4893–4908 (2008).

[2] Xinchen Wang, Siegfried Blechert, and Markus Antonietti, "Polymeric graphitic carbon nitride for heterogeneous photocatalysis," ACS Catalysis **2**, 1596–1606 (2012).

[3] Zaiwang Zhao, Yanjuan Sun, and Fan Dong, "Graphitic carbon nitride based nanocomposites: a review," Nanoscale **7**, 15–37 (2015).

[4] Jian Liu, Hongqiang Wang, and Markus Antonietti, "Graphitic carbon nitride "reloaded": emerging applications beyond (photo)catalysis," Chem. Soc. Rev. **45**, 2308–2326 (2016).

[5] Xinchen Wang, Kazuhiko Maeda, Arne Thomas, Kazuhiro Takanabe, Gang Xin, Johan M. Carlsson, Kazunari Domen, and Markus Antonietti, "A metal-free polymeric photocatalyst for hydrogen production from water under visible light," Nature Materials **8**, 76–80 (2009).

[6] Jonas Wirth, Rainer Neumann, Markus Antonietti, and Peter Saalfrank, "Adsorption and photocatalytic splitting of water on graphitic carbon nitride: a combined first principles and semiempirical study," Phys. Chem. Chem.

Phys. **16**, 15917–15926 (2014).
[7] Luis Miguel Azofra, Douglas R. MacFarlane, and Chenghua Sun, "A dft study of planar vs. corrugated graphene-like carbon nitride (g-c3n4) and its role in the catalytic performance of co2 conversion," Phys. Chem. Chem. Phys. **18**, 18507–18514 (2016).
[8] Jan Gerit Brandenburg, Andrea Zen, Martin Fitzner, Benjamin Ramberger, Georg Kresse, Theodoros Tsatsoulis, Andreas Grüneis, Angelos Michaelides, and Dario Alfè, "Physisorption of water on graphene: Subchemical accuracy from many-body electronic structure methods," The Journal of Physical Chemistry Letters **10**, 358–368 (2019), https://doi.org/10.1021/acs.jpclett.8b03679.
[9] Yasmine S. Al-Hamdani, Mariana Rossi, Dario Alfè, Theodoros Tsatsoulis, Benjamin Ramberger, Jan Gerit Brandenburg, Andrea Zen, Georg Kresse, Andreas Grüneis, Alexandre Tkatchenko, and Angelos Michaelides, "Properties of the water to boron nitride interaction: From zero to two dimensions with benchmark accuracy," The Journal of Chemical Physics **147**, 044710 (2017), https://doi.org/10.1063/1.4985878.
[10] Thomas Gruber, Ke Liao, Theodoros Tsatsoulis, Felix Hummel, and Andreas Grüneis, "Applying the coupled-cluster ansatz to solids and surfaces in the thermodynamic limit," Phys. Rev. X **8**, 021043 (2018).
[11] Tobias Schäfer, Florian Libisch, Georg Kresse, and Andreas Grüneis, "Local embedding of coupled cluster theory into the random phase approximation using plane waves," The Journal of Chemical Physics **154**, 011101 (2021), 2012.06165.
[12] B. G. Dick and A. W. Overhauser, "Theory of the dielectric constants of alkali halide crystals," Physical Review **112**, 90–103 (1958).
[13] J E Inglesfield, "A method of embedding," Journal of Physics C: Solid State Physics **14**, 3795–3806 (1981).
[14] Thom Vreven, Keiji Morokuma, Ödön Farkas, H. Bernhard Schlegel, and Michael J. Frisch, "Geometry optimization with QM/MM, ONIOM, and other combined methods. i. microiterations and constraints," Journal of Computational Chemistry **24**, 760–769 (2003).
[15] Frederick R. Manby, Martina Stella, Jason D. Goodpaster, and Thomas F. Miller, "A simple, exact density-functional-theory embedding scheme," Journal of Chemical Theory and Computation **8**, 2564–2568 (2012).
[16] "Nobel prizes 2013 m. karplus, m. levitt, a. warshel," Angewandte Chemie International Edition **52**, 11972–11972 (2013).
[17] Jonathan Nafziger and Adam Wasserman, "Density-based partitioning methods for ground-state molecular calculations," The Journal of Physical Chemistry A **118**, 7623–7639 (2014).
[18] F. Libisch, M. Marsman, J. Burgdörfer, and G. Kresse, "Embedding for bulk systems using localized atomic orbitals," The Journal of Chemical Physics **147**, 034110 (2017).
[19] Katharina Doblhoff-Dier, Geert-Jan Kroes, and Florian Libisch, "Density functional embedding for periodic and nonperiodic diffusion monte carlo calculations," Physical Review B **98** (2018), 10.1103/physrevb.98.085138.
[20] Brandon Eskridge, Henry Krakauer, and Shiwei Zhang, "Local embedding and effective downfolding in the auxiliary-field quantum monte carlo method," Journal of Chemical Theory and Computation **15**, 3949–3959 (2019), pMID: 31244125, https://doi.org/10.1021/acs.jctc.8b01244.
[21] Niklas Niemeyer, Johannes Tölle, and Johannes Neugebauer, "Approximate versus exact embedding for chiroptical properties: Reconsidering failures in potential and response," Journal of Chemical Theory and Computation **16**, 3104–3120 (2020).
[22] Niranjan Govind, Yan Alexander Wang, and Emily A. Carter, "Electronic-structure calculations by first-principles density-based embedding of explicitly correlated systems," The Journal of Chemical Physics **110**, 7677–7688 (1999), https://doi.org/10.1063/1.478679.
[23] Gerald Knizia and Garnet Kin-Lic Chan, "Density matrix embedding: A simple alternative to dynamical mean-field theory," Phys. Rev. Lett. **109**, 186404 (2012).
[24] Ireneusz W. Bulik, Weibing Chen, and Gustavo E. Scuseria, "Electron correlation in solids via density embedding theory," The Journal of Chemical Physics **141**, 054113 (2014), https://doi.org/10.1063/1.4891861.
[25] Jason D. Goodpaster, Taylor A. Barnes, Frederick R. Manby, and Thomas F. Miller, "Accurate and systematically improvable density functional theory embedding for correlated wavefunctions," The Journal of Chemical Physics **140**, 18A507 (2014).
[26] Tomasz A. Wesolowski, Sapana Shedge, and Xiuwen Zhou, "Frozen-density embedding strategy for multilevel simulations of electronic structure," Chemical Reviews **115**, 5891–5928 (2015).
[27] Oliver Masur, Martin Schütz, Lorenzo Maschio, and Denis Usvyat, "Fragment-Based Direct-Local-Ring-Coupled-Cluster Doubles Treatment Embedded in the Periodic Hartree–Fock Solution," Journal of Chemical Theory and Computation **12**, 5145–5156 (2016).
[28] Martin Schütz, Lorenzo Maschio, Antti J. Karttunen, and Denis Usvyat, "Exfoliation Energy of Black Phosphorus Revisited: A Coupled Cluster Benchmark," The Journal of Physical Chemistry Letters **8**, 1290–1294 (2017).
[29] Denis Usvyat, Lorenzo Maschio, and Martin Schütz, "Periodic and fragment models based on the local correlation approach," WIREs Computational Molecular Science **8**, 1–27 (2018).
[30] Dhabih V. Chulhai and Jason D. Goodpaster, "Projection-based correlated wave function in density functional theory embedding for periodic systems," Journal of Chemical Theory and Computation **14**, 1928–1942 (2018), pMID: 29494155, https://doi.org/10.1021/acs.jctc.7b01154.
[31] Sebastian J. R. Lee, Feizhi Ding, Frederick R. Manby, and Thomas F. Miller, "Analytical gradients for projection-based wavefunction-in-DFT embedding," The Journal of Chemical Physics **151**, 064112 (2019).
[32] Joachim Sauer, "Ab Initio Calculations for Molecule–Surface Interactions with Chemical Accuracy," Accounts of Chemical Research **52**, 3502–3510 (2019).
[33] Hung Hsuan Lin, Lorenzo Maschio, Daniel Kats, Denis Usvyat, and Thomas Heine, "Fragment-Based Restricted Active Space Configuration Interaction with Second-Order Corrections Embedded in Periodic Hartree-Fock Wave Function," Journal of Chemical Theory and Computation **16**, 7100–7108 (2020).
[34] Ji Chen, Nikolay A. Bogdanov, Denis Usvyat, Wei Fang, Angelos Michaelides, and Ali Alavi, "The color center singlet state of oxygen vacancies in TiO 2," The Journal of Chemical Physics **153**, 204704 (2020).

[35] Bryan T. G. Lau, Gerald Knizia, and Timothy C. Berkelbach, "Regional Embedding Enables High-Level Quantum Chemistry for Surface Science," The Journal of Physical Chemistry Letters **12**, 1104–1109 (2021), arXiv:2010.00527.

[36] Max Nusspickel and George H. Booth, "Systematic improvability in quantum embedding for real materials," (2021), arXiv:2107.04916.

[37] Merzuk Kaltak, Jiří Klimeš, and Georg Kresse, "Cubic scaling algorithm for the random phase approximation: Self-interstitials and vacancies in Si," Physical Review B - Condensed Matter and Materials Physics **90**, 054115 (2014).

[38] See Supplemental Material .

[39] G. Kresse and J. Furthmüller, "Efficiency of ab-initio total energy calculations for metals and semiconductors using a plane-wave basis set," Computational Materials Science **6**, 15–50 (1996).

[40] Andreas Grüneis, George H. Booth, Martijn Marsman, James Spencer, Ali Alavi, and Georg Kresse, "Natural orbitals for wave function based correlated calculations using a plane wave basis set," J. Chem. Theory Comput. **7**, 2780–2785 (2011).

[41] Benjamin Ramberger, Zoran Sukurma, Tobias Schäfer, and Georg Kresse, "RPA natural orbitals and their application to post-Hartree-Fock electronic structure methods," The Journal of Chemical Physics **151**, 214106 (2019), arXiv:1909.07089.

[42] M. Marsman, A. Grüneis, J. Paier, and G. Kresse, "Second-order Moller-Plesset perturbation theory applied to extended systems. I. Within the projector-augmented-wave formalism using a plane wave basis set," The Journal of Chemical Physics **130**, 184103 (2009).

[43] Tobias Schäfer, Benjamin Ramberger, and Georg Kresse, "Quartic scaling MP2 for solids: A highly parallelized algorithm in the plane wave basis," The Journal of Chemical Physics **146**, 104101 (2017).

[44] P. E. Blöchl, "Projector augmented-wave method," Physical Review B **50**, 17953–17979 (1994).

[45] Felix Hummel, Theodoros Tsatsoulis, and Andreas Grüneis, "Low rank factorization of the coulomb integrals for periodic coupled cluster theory," The Journal of Chemical Physics **146**, 124105 (2017), https://doi.org/10.1063/1.4977994.

[46] Traian Abrudan, Jan Eriksson, and Visa Koivunen, "Conjugate gradient algorithm for optimization under unitary matrix constraint," Signal Processing **89**, 1704–1714 (2009).

[47] Susi Lehtola and Hannes Jónsson, "Unitary Optimization of Localized Molecular Orbitals," Journal of Chemical Theory and Computation **9**, 5365–5372 (2013).

[48] Susi Lehtola and Hannes Jónsson, "Pipek-mezey orbital localization using various partial charge estimates," Journal of Chemical Theory and Computation **10**, 642–649 (2014).

[49] Elvar Ö. Jónsson, Susi Lehtola, Martti Puska, and Hannes Jónsson, "Theory and Applications of Generalized Pipek–Mezey Wannier Functions," Journal of Chemical Theory and Computation **13**, 460–474 (2017), 1608.06396.

[50] Neil W. N. David Mermin Ashcroft, *Solid-State Electronics*, Vol. 9 (Cengage Learning, Inc, 1966) pp. 939–942, arXiv:arXiv:1011.1669v3.

[51] János Pipek and Paul G. Mezey, "A fast intrinsic localization procedure applicable for a b i n i t i o and semiempirical linear combination of atomic orbital wave functions," The Journal of Chemical Physics **90**, 4916–4926 (1989).

[52] Gerald Knizia, "Intrinsic Atomic Orbitals: An Unbiased Bridge between Quantum Theory and Chemical Concepts," Journal of Chemical Theory and Computation **9**, 4834–4843 (2013), 1306.6884.

[53] C. Franchini, R. Kováčik, M. Marsman, S. Sathyanarayana Murthy, J. He, C. Ederer, and G. Kresse, "Maximally localized Wannier functions in LaMnO 3 within PBE + U , hybrid functionals and partially self-consistent GW: an efficient route to construct ab initio tight-binding parameters for e g perovskites," Journal of Physics: Condensed Matter **24**, 235602 (2012).

[54] Tomasz Janowski, "Near Equivalence of Intrinsic Atomic Orbitals and Quasiatomic Orbitals," Journal of Chemical Theory and Computation **10**, 3085–3091 (2014).

[55] Andreas Irmler, Alejandro Gallo, and Andreas Grüneis, "Focal-point approach with pair-specific cusp correction for coupled-cluster theory," The Journal of Chemical Physics **154**, 234103 (2021), arXiv:2103.06788.

[56] Andreas Irmler, Alejandro Gallo, Felix Hummel, and Andreas Grüneis, "Duality of ring and ladder diagrams and its importance for many-electron perturbation theories," Phys. Rev. Lett. **123**, 156401 (2019).

[57] Andreas Irmler and Andreas Grüneis, "Particle-particle ladder based basis-set corrections applied to atoms and molecules using coupled-cluster theory," The Journal of Chemical Physics **151**, 104107 (2019), https://doi.org/10.1063/1.5110885.

[58] W Kutzelnigg and JD Morgan, "Rates of convergence of the partial-wave expansions of atomic correlation energies," J. Chem. Phys. **96**, 4484–4508 (1992).

[59] Yu-ya Ohnishi and Seiichiro Ten-no, "Alternative formulation of explicitly correlated third-order moller–plesset perturbation theory," Mol. Phys. **111**, 2516–2522 (2013).

[60] Stefan Grimme, Stephan Ehrlich, and Lars Goerigk, "Effect of the damping function in dispersion corrected density functional theory," Journal of Computational Chemistry **32**, 1456–1465 (2011).

[61] Alexandre Tkatchenko and Matthias Scheffler, "Accurate molecular van der waals interactions from ground-state electron density and free-atom reference data," Phys. Rev. Lett. **102**, 073005 (2009).

[62] Ji ří Klimeš, David R. Bowler, and Angelos Michaelides, "Van der waals density functionals applied to solids," Phys. Rev. B **83**, 195131 (2011).

[63] Alberto Ambrosetti, Anthony M. Reilly, Robert A. DiStasio, and Alexandre Tkatchenko, "Long-range correlation energy calculated from coupled atomic response functions," The Journal of Chemical Physics **140**, 18A508 (2014).

[64] Andreas Grüneis, Martijn Marsman, and Georg Kresse, "Second-order Møller–Plesset perturbation theory applied to extended systems. II. Structural and energetic properties," The Journal of Chemical Physics **133**, 074107 (2010).

[65] David C Langreth and John P Perdew, "Exchange-correlation energy of a metallic surface: Wave-vector analysis," Physical Review B **15**, 2884–2901 (1977).

[66] Stefan Kurth and John P. Perdew, "Density-functional correction of random-phase-approximation correlation with results for jellium surface energies," Physical Re-

view B **59**, 10461–10468 (1999).

[67] Xinguo Ren, Patrick Rinke, and Matthias Scheffler, "Exploring the random phase approximation: Application to CO adsorbed on Cu(111)," Physical Review B **80**, 045402 (2009).

[68] Judith Harl, Laurids Schimka, and Georg Kresse, "Assessing the quality of the random phase approximation for lattice constants and atomization energies of solids," Physical Review B - Condensed Matter and Materials Physics **81** (2010), 10.1103/PhysRevB.81.115126.

[69] L Schimka, J Harl, A Stroppa, A. Grüneis, M Marsman, F Mittendorfer, and G Kresse, "Accurate surface and adsorption energies from many-body perturbation theory," Nature Materials **9**, 741–744 (2010).

[70] John F. Dobson and Tim Gould, "Calculation of dispersion energies," Journal of Physics: Condensed Matter **24**, 073201 (2012).

[71] Zidan Yan, John P Perdew, and Stefan Kurth, "Density functional for short-range correlation: Accuracy of the random-phase approximation for isoelectronic energy changes," Physical Review B **61**, 16430–16439 (2000).

[72] Andreas Grüneis, Martijn Marsman, Judith Harl, Laurids Schimka, and Georg Kresse, "Making the random phase approximation to electronic correlation accurate," The Journal of Chemical Physics **131**, 154115 (2009).

[73] Tobias Schäfer, Nathan Daelman, and Núria López, "Cerium Oxides without U : The Role of Many-Electron Correlation," The Journal of Physical Chemistry Letters **12**, 6277–6283 (2021).

# SUPPLEMENTARY MATERIAL FOR THE ARTICLE
# Surface science using coupled cluster theory via local Wannier functions and in-RPA-embedding: the case of water on graphitic carbon nitride

Tobias Schäfer,[1,*] Alejandro Gallo,[1] Andreas Irmler,[1] Felix Hummel,[1] and Andreas Grüneis[1]

[1] *Institute for Theoretical Physics, TU Wien, Wiedner Hauptstraße 8-10/136, A-1040 Vienna, Austria*

* tobias.schaefer@tuwien.ac.at

## CONVERGENCE OF THE ADSORPTION ENERGY FOR NON-RELAXED SUPERCELLS

Here we show an analog to Fig. 4 of the manuscript for the case of equivalent (i.e. non-relaxed) atomic positions in the layer in the supercells with and without water. To save computing time, a smaller plane-wave basis (`ENCUT` = 300 eV) and less natural orbitals (40 instead of 60) where chosen for the correlation energy calculations.

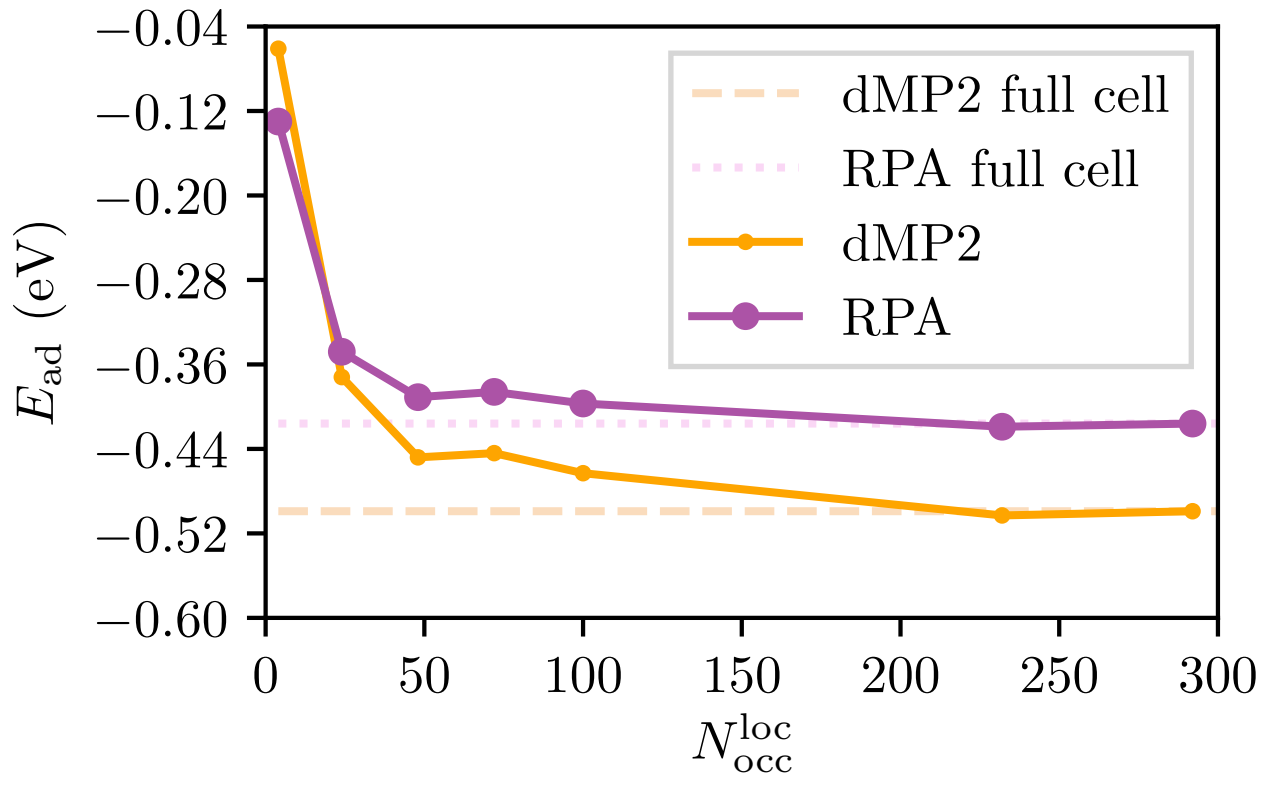

## ATOMIC STRUCTURES

In the following we list all `POSCAR` files as used in VASP to define the atomic species and positions. We also provide the Hartree-Fock energy obtained with the used PAW pseudo potential.

### $H_2O@gC_3N_4$

Hartree-Fock energy: −2098.534 eV.

```
H2O@gC3N4
   1.0000000000000000
  20.3700008391999994    0.0000000000000000    0.0000000000000000
  10.1849989889000003   17.6409373760999983    0.0000000000000000
   0.0000000000000000    0.0000000000000000   20.0000000000000000
   C    N    O    H
  54   72    1    2
Direct
   0.0496039867669697    0.1483298731768369    0.5564435091556922
   0.0458560036045063    0.2557866413184077    0.5216694554970678
   0.0459684904918915    0.3752252555712546    0.5216584454609722
   0.3826451796812914    0.4814272067640431    0.5563574140017337
   0.0498096707403496    0.4789746408757268    0.4875157241297723
   0.3779221075782361    0.5887568994500983    0.5207181059538137
   0.0508283010029643    0.5889995851456652    0.4828899925450301
   0.1490220132452628    0.3795613891262736    0.5570204935267483
   0.1574797762628455    0.4888182210905045    0.5238809398318888
   0.2581788871805645    0.3819126772790122    0.5639720544895737
   0.0390718323916386    0.7152972465518133    0.4827202437248444
   0.3782057412787609    0.7081068104596497    0.5219392475781964
   0.7153198519972863    0.8152647360006772    0.5571006547472471
   0.0333318685396401    0.8217032672809398    0.5232740393781254
   0.3823560722813698    0.8121857901677347    0.4884538821898156
   0.7115676607241209    0.9230241748975986    0.5231367574571041
   0.0401392456135033    0.9219558689254349    0.5635289583264974
   0.3848821590250230    0.9214706809590695    0.4826698188192755
```

| | | |
|---|---|---|
| 0.1506684620786370 | 0.7130910483772197 | 0.4879052091322137 |
| 0.4821349884836069 | 0.7120653389978451 | 0.5567257670932257 |
| 0.1513983123323246 | 0.8123643388369490 | 0.5568597062330257 |
| 0.4915601263951209 | 0.8204638203385343 | 0.5227200511673384 |
| 0.2586549178445293 | 0.7086752747715002 | 0.5217115523258234 |
| 0.5917318312767588 | 0.7137718363921313 | 0.5638556808059100 |
| 0.0513490391636347 | 0.0369252544817911 | 0.5627571297809066 |
| 0.3736432157563244 | 0.0477697607799498 | 0.4830974666018089 |
| 0.7119579907078487 | 0.0423288040669866 | 0.5234974886544188 |
| 0.3669661621244152 | 0.1545633397055781 | 0.5238298324138918 |
| 0.7160562125581863 | 0.1457877184186257 | 0.4886341982270235 |
| 0.3728641298964249 | 0.2553734214796173 | 0.5640120986514794 |
| 0.7181725429042489 | 0.2551244465750589 | 0.4826284508971239 |
| 0.1578797077351390 | 0.0306284943190058 | 0.5218612411741400 |
| 0.4847714650259187 | 0.0465080607947084 | 0.4884992736106204 |
| 0.8156624046553175 | 0.0463155980435806 | 0.5578068327833073 |
| 0.1493742300600266 | 0.1480449559028373 | 0.4872819559576317 |
| 0.4845553972542402 | 0.1461248156224580 | 0.5576570112675482 |
| 0.8250894525231652 | 0.1544293762658469 | 0.5223363571578854 |
| 0.2587818938209949 | 0.0364954356794479 | 0.4820568603858473 |
| 0.5922973590314305 | 0.0426196989365717 | 0.5231555964459916 |
| 0.9252922620452554 | 0.0480754622382189 | 0.5632768344036030 |
| 0.3846355202421340 | 0.3700753698652107 | 0.5631170314168130 |
| 0.7071062597883162 | 0.3806066291807504 | 0.4859208066739656 |
| 0.7018775097577560 | 0.4861597392728051 | 0.5288838380667416 |
| 0.7075643903394033 | 0.5879053693556340 | 0.5668373695197277 |
| 0.4910077449298848 | 0.3638717975972766 | 0.5221365097091154 |
| 0.8184180643132096 | 0.3794958593849085 | 0.4902795912114781 |
| 0.4819135384044497 | 0.4813365343943217 | 0.4873425102334575 |
| 0.8196035101717778 | 0.4790435362116349 | 0.5587753646579539 |
| 0.5917797639670703 | 0.3701846404286236 | 0.4829423132682341 |
| 0.9266038085141965 | 0.3753663117437859 | 0.5229737293678520 |
| 0.7173999656927393 | 0.7036721895245733 | 0.5633283037761337 |
| 0.8234489440491087 | 0.6976579502096153 | 0.5215157648716152 |
| 0.8151253283837814 | 0.8153943292119824 | 0.4884382298991756 |
| 0.9243517741233506 | 0.7037477294190598 | 0.4820108300285549 |
| 0.0206430193357825 | 0.1086381083259750 | 0.5840429235323504 |
| 0.0154165038801805 | 0.2237129460986270 | 0.5595487163863884 |
| 0.1085074952394543 | 0.2234629009045117 | 0.4839476026930036 |
| 0.0060694043165961 | 0.3354318184477893 | 0.5217280330992797 |
| 0.3538516747777440 | 0.4417095337991556 | 0.5842061079119050 |
| 0.0157301297454579 | 0.4377056709074169 | 0.4837325527275481 |
| 0.3480242364390894 | 0.5568116695015143 | 0.5592282788179261 |
| 0.0206683245964106 | 0.5479471360844762 | 0.4605052310572391 |
| 0.1082679695002748 | 0.3448797165259609 | 0.5596552282376248 |
| 0.1185496147996325 | 0.4487788522246098 | 0.5225978130278640 |
| 0.4399633312054445 | 0.5566565721288811 | 0.4823860750719167 |
| 0.1182600480228522 | 0.5631119621464681 | 0.5120523684415856 |
| 0.2174963922198233 | 0.3506581525017248 | 0.5847143975577869 |
| 0.2314576447338031 | 0.4504626039463011 | 0.5367989476431863 |
| 0.0046382978017357 | 0.6693631534701201 | 0.4764987429351480 |
| 0.3383013116254293 | 0.6686263403972903 | 0.5212597407663034 |
| 0.6867188224710279 | 0.7751996573664188 | 0.5848273281498543 |
| 0.3475739094027841 | 0.7715742240459388 | 0.4854457879349793 |
| 0.6807509527914368 | 0.8906187301129723 | 0.5602113810955411 |
| 0.3535589485094480 | 0.8809909986946834 | 0.4613507741069643 |
| 0.1190390241135033 | 0.9953412358227494 | 0.5341232329679442 |
| 0.1105331327761728 | 0.6843221631238213 | 0.4608336888555072 |
| 0.4415458545595363 | 0.6771160820518924 | 0.5592053779179845 |
| 0.1123165792883917 | 0.7821634824231586 | 0.5225624226192681 |
| 0.4518158779799712 | 0.7812026883199205 | 0.5225404375898282 |
| 0.7745743069718070 | 0.8908273703575054 | 0.4859240679694445 |
| 0.1120256307303033 | 0.8811357053721344 | 0.5842838865201274 |
| 0.4531446947657772 | 0.8946085862434092 | 0.5105412360657147 |
| 0.2320920352869405 | 0.9950680043973964 | 0.5099958509603408 |
| 0.2260188750777571 | 0.6784063330682178 | 0.4843416392408664 |
| 0.5505632609627045 | 0.6830743153652492 | 0.5847034588565496 |
| 0.2268126571291351 | 0.7712123368770695 | 0.5595796146831247 |
| 0.5657100535567664 | 0.7814436349047844 | 0.5351960095585958 |
| 0.0056731413667284 | 0.0023093807572724 | 0.5698850630360163 |
| 0.3392506123851256 | 0.0017830343936393 | 0.4764754875670404 |
| 0.6719640839341313 | 0.0026227042726610 | 0.5233956782870651 |
| 0.3318819569541180 | 0.1154040048184002 | 0.5118372316287024 |
| 0.6814004981058233 | 0.1050489712861292 | 0.4859280479200737 |
| 0.3312295403248132 | 0.2285957931936555 | 0.5365722517246220 |
| 0.6870607836554065 | 0.2145745666762896 | 0.4612783715857630 |
| 0.4524305413950714 | 0.3283171258910547 | 0.5346772509822715 |
| 0.4451065171033580 | 0.0172406663340694 | 0.4613107219099339 |
| 0.7746931722194663 | 0.0118852237252093 | 0.5611125386418511 |
| 0.1186058460283309 | 0.1095610206887608 | 0.5217167688712885 |
| 0.4459599977980733 | 0.1155687140507707 | 0.5230661951805924 |
| 0.7853669071039073 | 0.1151861461480966 | 0.5228912409390060 |
| 0.4447012917616595 | 0.2148336090412579 | 0.5849866044455863 |
| 0.7865464488391971 | 0.2284843357382412 | 0.5100777878576522 |
| 0.5652685038406317 | 0.3284588256703405 | 0.5103608733447922 |
| 0.5601435873900314 | 0.0122214469822271 | 0.4852814136728068 |
| 0.8843931012597610 | 0.0175427510043013 | 0.5851979554055277 |
| 0.2183922835795216 | 0.1080850866026346 | 0.4603428291374665 |
| 0.5599791079377905 | 0.1052580883811533 | 0.5608259970408428 |
| 0.8993003425734901 | 0.1155322575277082 | 0.5342029645965805 |
| 0.3384909588653559 | 0.3357637857579341 | 0.5700330706088205 |
| 0.6723767207221134 | 0.3354800280371955 | 0.4773778065042970 |
| 0.6667221520890424 | 0.4466194769528252 | 0.5180626557664825 |
| 0.6657103227370076 | 0.5598002737727932 | 0.5422450391562039 |
| 0.7844194251773977 | 0.6624248402893057 | 0.5335888040396080 |
| 0.7777198370409196 | 0.3515799602112147 | 0.4626404589297528 |
| 0.4512913633799742 | 0.4427221807191022 | 0.5212728054803298 |
| 0.7804795589759898 | 0.4479523945630100 | 0.5258679039399958 |
| 0.7801820741686498 | 0.5480567632920467 | 0.5857926412711820 |
| 0.8974931165989842 | 0.6619532493906898 | 0.5090535245180258 |
| 0.8935903900808311 | 0.3449643259287368 | 0.4861679549909456 |
| 0.5515492753363801 | 0.4417703164490480 | 0.4613043999623850 |
| 0.8950607101599757 | 0.4383214173014001 | 0.5603852240764031 |
| 0.6720985226370091 | 0.6684203002155988 | 0.5714673718174543 |
| 0.9977170061862525 | 0.7830620699330408 | 0.5112869561786606 |
| 0.9980998729458490 | 0.8957134872062283 | 0.5359641781265382 |
| 0.7843279191396387 | 0.7766137310950847 | 0.5223468380949347 |
| 0.8839251631415579 | 0.7756211762634050 | 0.4611134084212349 |
| 0.5359642740522966 | 0.5060662183470885 | 0.6130439798945234 |
| 0.5097604480578262 | 0.5612556917352064 | 0.6075760764984104 |
| 0.5799888927952579 | 0.4882861014483103 | 0.5832691482637062 |

**$gC_3N_4$**

Hartree-Fock energy: $-2069.042$ eV.

```
gC3N4
  1.0000000000000000
 20.3700008391999994   0.0000000000000000   0.0000000000000000
 10.1849989889000003  17.6409373760999983   0.0000000000000000
  0.0000000000000000   0.0000000000000000  20.0000000000000000
  C    N
 54   72
Direct
  0.0488303140716052  0.1478536857864759  0.5576562332472806
  0.0449130518337105  0.2554147638846518  0.5230533878147474
  0.0448097201085649  0.3749595439865857  0.5231426268223338
  0.3821636474049409  0.4811870191198117  0.5576562332472806
  0.0487174838278628  0.4787101638249346  0.4887545683081748
  0.3782463851670462  0.5887480972179805  0.5230533878147474
  0.0508615811712641  0.5881218265897876  0.4826150806734776
  0.1486372768070802  0.3788042704963207  0.5573267787430988
  0.1575963625046865  0.4875078762657816  0.5228826333422886
  0.2581311748795082  0.3808540493358762  0.5633470654273818
  0.0396041236610879  0.7143496021578034  0.4820709198402200
  0.3781430534418936  0.7082928773199213  0.5231426268223338
  0.7154969807382694  0.8145203524531471  0.5576562332472806
  0.0334593752541961  0.8212215802028563  0.5220000603945452
  0.3820508171611915  0.8120434971582633  0.4887545683081748
  0.7115797185003749  0.9220814305513162  0.5230533878147474
  0.0396922448974063  0.9214149251213993  0.5631283016058377
  0.3841949145045927  0.9214551599231233  0.4826150806734776
  0.1511634194669931  0.7121268514285490  0.4883981427924712
  0.4819706101404087  0.7121376038296491  0.5573267787430988
  0.1510208070372490  0.8117460331891944  0.5575203718767975
  0.4909296958380223  0.8208412095991173  0.5228826333422886
  0.2587395321778988  0.7081193165473397  0.5231677269445588
  0.5914645082128437  0.7141873826692049  0.5633470654273818
  0.0509939941043123  0.0362971741836393  0.5640543668483713
  0.3729374569944237  0.0476829354911319  0.4820709198402200
  0.7114763867752293  0.0416262106532570  0.5231426268223338
  0.3667927085875318  0.1545549135361848  0.5220000603945452
  0.7153841504945270  0.1453768304915989  0.4887545683081748
  0.3730255782307419  0.2547482584547349  0.5631283016058377
  0.7175282478379285  0.2547884932564591  0.4826150806734776
  0.1578226974787084  0.0301287301495509  0.5239405976700837
  0.4844967528003288  0.0454601847618776  0.4883981427924712
  0.8153039434737444  0.0454709371629849  0.5573267787430988
  0.1484399264723505  0.1476951974980128  0.4885048801718201
  0.4843541403705776  0.1450793665225230  0.5575203718767975
  0.8242630291713580  0.1541745429324531  0.5228826333422886
  0.2580460131065728  0.0363575031681650  0.4828453008761566
  0.5920728655112345  0.0414526498806755  0.5231677269445588
  0.9247978415461794  0.0475207160025405  0.5633470654273818
  0.3843273274376409  0.3696305075169750  0.5640543668483713
  0.7062707903277523  0.3810162688244675  0.4820709198402200
  0.7001260419208675  0.4878882468695205  0.5220000603945452
  0.7063589115640775  0.5880815917880707  0.5631283016058377
  0.4911560308120442  0.3634620634828866  0.5239405976700837
  0.8178300861336643  0.3787935180952133  0.4883981427924712
  0.4817732598056862  0.4810285308313412  0.4885048801718201
  0.8176874737039134  0.4784126998558588  0.5575203718767975
  0.5913793464399086  0.3696908365014936  0.4828453008761566
  0.9254061988445631  0.3747859832140041  0.5231677269445588
```

| | | |
|---|---|---|
| 0.7176606607709765 | 0.7029638408503035 | 0.5640543668483713 |
| 0.8244893641453728 | 0.6967953968162154 | 0.5239405976700837 |
| 0.8151065931390217 | 0.8143618641646774 | 0.4885048801718201 |
| 0.9247126797732443 | 0.7030241698348292 | 0.4828453008761566 |
| 0.0196958429382165 | 0.1081784758588351 | 0.5849190034168025 |
| 0.0144120318179337 | 0.2232506120098144 | 0.5607175594375987 |
| 0.1076004569526298 | 0.2230930314683252 | 0.4853943036805292 |
| 0.0050686964521933 | 0.3350791825128605 | 0.5229982733944822 |
| 0.3530291762715522 | 0.4415118091921708 | 0.5849190034168025 |
| 0.0140262361459642 | 0.4380007304508958 | 0.4859443375703842 |
| 0.3477453651512693 | 0.5565839453431503 | 0.5607175594375987 |
| 0.0197594775845619 | 0.5474959861025305 | 0.4613741611297458 |
| 0.1077946852527338 | 0.3442488871651824 | 0.5604332118028591 |
| 0.1180934240588844 | 0.4480368001006907 | 0.5229164004555472 |
| 0.4409337902859654 | 0.5564263648016610 | 0.4853943036805292 |
| 0.1188890889012494 | 0.5616759524178259 | 0.5107166849205762 |
| 0.2174670755455373 | 0.3498014771515387 | 0.5845841465742462 |
| 0.2317685298664970 | 0.4487853659656236 | 0.5350845888363301 |
| 0.0050742735401237 | 0.6685059199347421 | 0.4761763822642165 |
| 0.3384020297855219 | 0.6684125158461961 | 0.5229982733944822 |
| 0.6863625096048807 | 0.7748451425254992 | 0.5849190034168025 |
| 0.3473595694793000 | 0.7713340637842316 | 0.4859443375703842 |
| 0.6810786984845980 | 0.8899172786764860 | 0.5607175594375987 |
| 0.3530928109178905 | 0.8808293194358662 | 0.4613741611297458 |
| 0.1194124267732747 | 0.9945199579083777 | 0.5367127083448849 |
| 0.1114126825961909 | 0.6831176947871417 | 0.4609963525474581 |
| 0.4411280185860623 | 0.6775822204985109 | 0.5604332118028591 |
| 0.1124204340743875 | 0.7814640356055002 | 0.5226093189341575 |
| 0.4514267573922131 | 0.7813701334340195 | 0.5229164004555472 |
| 0.7742671236193011 | 0.8897596981349896 | 0.4853943036805292 |
| 0.1111212268345796 | 0.8804655987392652 | 0.5849680576141100 |
| 0.4522224222345851 | 0.8950092857511616 | 0.5107166849205762 |
| 0.2319976238395033 | 0.9948743274931030 | 0.5118040433389373 |
| 0.2265723418112325 | 0.6775239961117164 | 0.4856306206144567 |
| 0.5508004088788732 | 0.6831348104848742 | 0.5845841465742462 |
| 0.2264080947958670 | 0.7707960362385551 | 0.5608527477277152 |
| 0.5651018631998252 | 0.7821186992989524 | 0.5350845888363301 |
| 0.0051691970358705 | 0.0017533503708866 | 0.5700702913475476 |
| 0.3384076068734522 | 0.0018392532680706 | 0.4761763822642165 |
| 0.6717353631188577 | 0.0017458491795319 | 0.5229982733944822 |
| 0.3312243838913563 | 0.1160276105265442 | 0.5095994515595306 |
| 0.6806929028126286 | 0.1046673971175673 | 0.4859443375703842 |
| 0.3315522937705745 | 0.2287584959082999 | 0.5340368903950868 |
| 0.6864261442512262 | 0.2141626527692019 | 0.4613741611297458 |
| 0.4527457601066033 | 0.3278532912417063 | 0.5367127083448849 |
| 0.4447460159295267 | 0.0164510281204704 | 0.4609963525474581 |
| 0.7744613519193978 | 0.0109155538318467 | 0.5604332118028591 |
| 0.1180385482216639 | 0.1090932943372465 | 0.5232072140837213 |
| 0.4457537674077163 | 0.1147973689388357 | 0.5226093189341575 |
| 0.7847600907255488 | 0.1147034667673551 | 0.5229164004555472 |
| 0.4444545601679084 | 0.2137989320726007 | 0.5849680576141100 |
| 0.7855557555679140 | 0.2283426190844975 | 0.5107166849205762 |
| 0.5653309571728321 | 0.3282076608264387 | 0.5118040433389373 |
| 0.5599056751445615 | 0.0108573294450451 | 0.4856306206144567 |
| 0.8841337422122018 | 0.0164681438182101 | 0.5845841465742462 |
| 0.2171562815095514 | 0.1077999136564738 | 0.4610012066094455 |
| 0.5597414281292030 | 0.1041293695718836 | 0.5608527477277152 |
| 0.8984351965331610 | 0.1154520326322879 | 0.5350845888363301 |

```
   0.3385025303692061   0.3350866837042222   0.5700702913475476
   0.6717409402067881   0.3351725866014063   0.4761763822642165
   0.6645577172246920   0.4493609438598798   0.5095994515595306
   0.6648856271039030   0.5620918292416355   0.5340368903950868
   0.7860790934399391   0.6611866245750420   0.5367127083448849
   0.7780793492628625   0.3497843614538060   0.4609963525474581
   0.4513718815549997   0.4424266276705822   0.5232072140837213
   0.7790871007410518   0.4481307022721644   0.5226093189341575
   0.7777878935012438   0.5471322654059366   0.5849680576141100
   0.8986642905061678   0.6615409941597673   0.5118040433389373
   0.8932390084778972   0.3441906627783808   0.4856306206144567
   0.5504896148428874   0.4411332469898023   0.4610012066094455
   0.8930747614625387   0.4374627029052193   0.5608527477277152
   0.6718358637025348   0.6684200170375582   0.5700702913475476
   0.9978910505580206   0.7826942771932086   0.5095994515595306
   0.9982189604372387   0.8954251625749712   0.5340368903950868
   0.7847052148883353   0.7757599610039108   0.5232072140837213
   0.8838229481762231   0.7744665803231383   0.4610012066094455
```

## $H_2O$

Hartree-Fock energy: $-29.286$ eV.

```
H2O
   1.0000000000000000
  20.3700008391999994   0.0000000000000000   0.0000000000000000
  10.1849989889000003  17.6409373760999983   0.0000000000000000
   0.0000000000000000   0.0000000000000000  20.0000000000000000
   O    H
   1    2
 Direct
   0.1941134177990449   0.2661924938318947   0.5404622682889695
   0.2465147797639702   0.2301969514217674   0.5290342598181113
   0.1643718024369889   0.2536105547463308   0.5105034718929176
```